\documentclass[11pt,a4paper]{article}
\usepackage[T1]{fontenc}
\usepackage[utf8]{inputenc}
\usepackage{amsmath}
\usepackage{amsfonts, amssymb, graphicx, authblk, geometry, color, xcolor, url}
\usepackage{jheppub}
\usepackage{cleveref}
\usepackage{graphicx}
\usepackage{caption}
\usepackage{subcaption}
\usepackage{float}
\usepackage{soul}
\usepackage{subcaption}
\usepackage{slashed}

\newcommand{\meff}{M_{\text{eff}}}

\newcommand{\tchi}{\tilde{\chi}}
\newcommand{\deltchi}{\delta\tilde{\chi}}
\newcommand{\tlog}{\text{log}\,}
\newcommand{\bgk}[1]{\bar{g}_{#1}(\eta)\,}

\newcommand{\plus}{\, + \,}

\newcommand{\dphiv}{\dot{\phi}_{0}}
\newcommand{\intl}[2]{\int_{#1}^{#2}}
\newcommand{\fnl}{f_{\text{NL}}}
\def\<{\langle\,}
\def\>{\,\rangle}

\newcommand{\Fig}[1]{figure~\ref{#1}}
\newcommand{\Eq}[1]{Eq.~(\ref{#1})}

\graphicspath{{figures/}}

\title{The scalar chemical potential in 
cosmological collider physics}
\author[a]{Arushi Bodas,}
\author[a,b,c]{Soubhik Kumar}
\author[a]{and Raman Sundrum}
\affiliation[a]{Maryland Center for Fundamental Physics, Department of Physics,\\University of Maryland, College Park, MD 20742, USA}
\affiliation[b]{Berkeley Center for Theoretical Physics, Department of Physics,
University of California, Berkeley, CA 94720, USA}
\affiliation[c]{Theoretical Physics Group, Lawrence Berkeley National Laboratory, Berkeley, CA 94720, USA}
\emailAdd{arushib@terpmail.umd.edu }
\emailAdd{soubhik@terpmail.umd.edu}
\emailAdd{raman@umd.edu}

\abstract{
Non-analyticity in co-moving momenta within the
non-Gaussian bispectrum is a distinctive sign of \emph{on-shell} particle production during inflation, presenting a unique opportunity for the ``direct detection'' of particles with masses as large as the inflationary Hubble scale ($H$).
However, the strength of such non-analyticity ordinarily drops exponentially by a Boltzmann-like factor as masses exceed $H$.
In this paper, we study an exception provided by a dimension-5 derivative coupling of the inflaton to heavy-particle currents, applying it specifically to the case of two real scalars.
The operator has a ``chemical potential'' form, which harnesses the large kinetic energy scale of the inflaton, $\dphiv^{1/2} \approx 60H$, to  act as an efficient source of scalar particle production.
Derivative couplings of inflaton ensure radiative stability of the slow-roll potential, which in turn maintains (approximate) scale-invariance of the inflationary correlations. 
We show that a signal not suffering Boltzmann suppression can be obtained in the bispectrum with strength $\fnl \sim \mathcal{O}(0.01-10)$ for an extended range of scalar masses $ \lesssim \dphiv^{1/2}$,  potentially as high as $10^{15}$ GeV, 
within the sensitivity of upcoming LSS and more futuristic 21-cm experiments. The mechanism does not invoke any particular fine-tuning of parameters or breakdown of perturbation-theoretic control.
The leading contribution appears at \emph{tree-level}, 
which makes the calculation analytically tractable and removes the loop-suppression as compared to earlier chemical potential studies of non-zero spins. 
The steady particle production allows us to infer the effective mass of the heavy particles and the chemical potential from the variation in bispectrum oscillations as a function of co-moving momenta. Our analysis sets the stage for generalization to heavy bosons with non-zero spin.
}
\begin{document}
\hspace{30em} UMD-PP-020-09
\maketitle
\flushbottom

\section{Introduction}

The paradigm of cosmic inflation (for a review see \cite{Baumann:2009ds}) gives a robust mechanism for the high degree of homogeneity and isotropy of the universe on very large scales, as the result of exponential spacetime expansion driven by classical inflaton scalar dynamics coupled to General Relativity.
Furthermore, quantum fluctuations in this paradigm lead to a structure of tiny inhomogeneities, beautifully consistent with what is observed, for example, in the Cosmic Microwave Background (CMB).
In particular, data suggests that these primordial fluctuations are scale-invariant, adiabatic and Gaussian to a good precision \cite{Akrami:2018odb,Akrami:2019izv}, supporting their origins from a single weakly coupled quantum field, minimally identified with the inflaton itself.
In general, interactions can give rise to small non-trivial $n$-point correlations of the inflaton field, which translates into non-Gaussianity (NG) of the primordial curvature fluctuations (for reviews see \cite{Bartolo:2004if, Chen:2010xka}).
But with the expected improvement in experimental precision in the near future, especially with Large-Scale Structure (LSS) data \cite{Alvarez:2014vva,2010AdAst2010E..64V}, we can hope to detect small NG, revealing subdominant interactions of the inflaton with itself or other fields. In this way, the properties of these other fields during inflation may be encoded in primordial NG observables.\footnote{This is true provided the NG induced by gravitational interactions in the late-time Universe can be modeled accurately and separated out. Keeping that in mind, in what follows we will focus exclusively on \textit{primordial} NG sourced during the inflationary era.
}

In more detail, the time-dependent inflationary space-time provides energy of order the expansion rate, i.e., the Hubble scale ($H$), that leads to the \emph{on-shell} production of any particle with mass $M\sim H$.
The particle production here is associated with the fact that a geodesic observer in a de Sitter (dS) space-time sees a \emph{thermal distribution} of particles at the Hawking temperature $T_{\text{Hawking}}= H/2\pi$ (see e.g.~\cite{Birrell:1982ix}). If these particles interact with the inflaton, once produced they can later decay into inflaton fluctuations, resulting in primordial NG \cite{Chen:2009zp}.
The cosmological production and later decay into inflatons of heavy particles leaves a unique non-local feature in the primordial 3-point function of curvature fluctuations ($\mathcal{R}$), also known as the bispectrum. When expressed in co-moving momentum space, this spacetime non-locality appears as momentum non-analyticity in
the so-called squeezed limit, where one of the momenta is soft, $k_1\sim k_2 \gg k_3$. Typically \cite{Chen:2009zp, Arkani-Hamed:2015bza, Chen:2012ge},
\begin{align}\label{eq:bisprectrum_general}
   \frac{\< {\cal R}_{\vec{k}_1} {\cal R}_{\vec{k}_2} {\cal R}_{\vec{k}_3}\>}{\<{\cal R} {\cal R}\>_{k_1}\<{\cal R} {\cal R}\>_{k_3}}   \propto e^{-\pi \mu}\, \left( \frac{k_3}{k_1} \right)^{3/2\pm i\mu} \, P_{s}(\cos \theta) +\cdots
    \qquad \text{where} \qquad \mu = \sqrt{\frac{M^2}{H^2}-\frac{9}{4}}.
\end{align}
Here $s$ is the spin of the heavy particle and $\theta$ is the angle between the hard and the soft momenta. We  see that the non-analytic dependence on the momentum ratio $k_1/k_3$, which is either oscillatory ($M > 3H/2 $) or non-standard power-law ($M \lesssim 3H/2$), and the angular dependence can be used to do \textit{spectroscopy} of the mass and the spin of the heavy particle. 
This opens up a new window into particle physics at energy scales as high as the inflationary Hubble scale ($\lesssim 5\times 10^{13}$ GeV \cite{Akrami:2018odb}), potentially orders of magnitude beyond the reach of any terrestrial collider. This program of research and its applications have been dubbed ``cosmological collider physics'' \cite{Chen:2009zp, Baumann:2011nk, Noumi:2012vr, Chen:2012ge, Pi:2012gf, Assassi:2012zq, Arkani-Hamed:2015bza, Dimastrogiovanni:2015pla, Chen:2016nrs, Lee:2016vti, Chen:2016uwp, Chen:2016hrz, Chen:2017ryl, Kehagias:2017cym, An:2017hlx, Kumar:2017ecc, Baumann:2017jvh, Chen:2018xck, Arkani-Hamed:2018kmz, Kumar:2018jxz, Wu:2019ohx, Li:2019ves, Alexander:2019vtb, Lu:2019tjj, Hook:2019zxa, Hook:2019vcn, Kumar:2019ebj, Liu:2019fag, Wang:2019gbi, Baumann:2019oyu, Li:2020xwr, Wang:2020uic, Wang:2020ioa, Testa:2020hox, Baumann:2020dch}.

While this is a very exciting program, a significant hurdle to fully exploit the reach of the ``cosmological collider'' is that the non-analytic signal gets exponentially smaller ($\sim e^{-\pi M/H}$) for $M\gg H$ as seen in \cref{eq:bisprectrum_general}. Intuitively, this is a ``Boltzmann suppression'' factor  due to the thermal-like nature of particle production at the Hawking temperature $T_{\text{Hawking}}=H/2\pi$, where the bispectrum gives the amplitude for particle production, $\propto e^{-M/2T_{\text{Hawking}}}$, corresponding to the standard Boltzmann probability $e^{-M/T_{\text{Hawking}}}$. On the other hand, for small masses $M \ll H$, we see that the non-analytic factor in $k_1/k_3$ becomes approximately analytic and difficult to disentangle from other NG sources, such as inflaton self-interactions. Therefore apparently, these two considerations very strongly restrict observable masses to a window of $\sim H$.

Of course, in addition to the production amplitude for new heavy particles, the strength of NG depends on the couplings to the inflaton, which we now consider. Radiative stability of the inflaton potential strongly suggests that it should have derivative couplings predominantly, schematically of the form
\begin{align}\label{eq:EFT_operators}
    \frac{1}{\Lambda^{2n+dim(\mathcal{O}_{\rm heavy})-4}} (\partial \phi)^n \mathcal{O}_{\rm heavy},
\end{align}
where ${\mathcal O}_{\rm heavy}$ is made from the heavy fields and  $\Lambda$ is the scale at which the non-renormalizable EFT description breaks down. While the exact cutoff depends on the assumptions about the nature of the UV-physics, plausibly it should be at least as big as the scale of inflaton kinetic energy in slow-roll inflation \cite{Creminelli:2003iq}, 
\begin{align}\label{eq:Lambda_cutoff}
    \Lambda > \sqrt{\dphiv} \gtrsim 60H,
\end{align}
where $\phi_0(t)$ denotes the classical inflaton field trajectory, and 
where we have used the fact that the CMB temperature power spectrum amplitude implies $H^4/\dphiv^2 \approx 10^{-7} $ \cite{Akrami:2018odb}. 
It therefore seems that the typical (dimensionless) couplings cannot offset Boltzmann suppression, since they 
are at best $\lesssim {\cal O}(1)$ when some of the inflaton fields are evaluated at the classical expectations $\dot{\phi}_0$, or are significantly suppressed.

The case $n=1$ in \cref{eq:EFT_operators} is exceptional however, with the inflaton coupling to a current made of heavy fields, $\partial_{\mu} \phi J^{\mu}_{\rm heavy}/ \Lambda$. 
This has been studied in the context of heavy fermionic matter \cite{Adshead:2015kza, Adshead:2018oaa, Chen:2018xck, Hook:2019zxa, Wang:2019gbi, Hook:2019vcn} and 
heavy spin-1 bosons \cite{Garretson:1992vt, Barnaby:2010vf, Wang:2019gbi, Wang:2020ioa}.
Here we will study the simple but interesting case of heavy scalar matter,
and show that there are dramatic new features enhancing the mass reach of cosmological collider physics in a controlled and robust manner.
The case of a single real heavy scalar, $\sigma$, is however trivial since the unique coupling is given by
\begin{align}\label{eq:phi_sigma_coupling}
    {\cal L}_{\rm int} \supset  \, \frac{g^{\mu \nu}(\partial_{\mu}\phi) (\partial_{\nu} \sigma) \sigma}{\Lambda}.
\end{align}
Integrating by parts and replacing the resulting 
$\Box \phi$ by $-V'(\phi)$ using the inflaton equation of motion, 
this only makes a mass contribution to $\sigma$ modulo small slow-roll corrections.  For the case of two real scalars, $\sigma_1,\, \sigma_2$, on the other hand, \cref{eq:phi_sigma_coupling} generalizes to
\begin{align}\label{eq:twoScalar_chemPot}
    {\cal L}_{\rm int} \supset  \, \left(\frac{1 }{\Lambda_1} g^{\mu \nu} (\partial_{\mu}\phi) (\partial_{\nu} \sigma_1) \sigma_2 - \frac{1 }{\Lambda_2}g^{\mu \nu}(\partial_{\mu}\phi) (\partial_{\nu} \sigma_2) \sigma_1\right).
\end{align}
For generic values of $\Lambda_1$ and $\Lambda_2$, these operators can not be eliminated by integration by parts. Inserting the time-dependent inflaton VEV generates a quadratic mixing between the two scalars,
\begin{align}\label{eq:scalar_mix_chemPot}
   \lambda_1 \dot{\sigma}_1  \sigma_2 - \lambda_2 \dot{\sigma}_2  \sigma_1,
\end{align}
where $\lambda_1 = \dphiv/\Lambda_1$ and $\lambda_2 =  \dphiv/\Lambda_2$ are constant in time up to slow-roll corrections. 
This form is reminiscent of the chemical potential for a complex scalar $\chi= \sigma_1 + i \sigma_2$.

Indeed, for the special case $\lambda_2 = \lambda_1 =\lambda$ and equal masses $M_1=M_2$, \cref{eq:scalar_mix_chemPot}
is the chemical potential $\sim \lambda J^0$, where
$J$ is the Noether current associated to $\chi$ phase rotations, $J^0 = \dot{\sigma}_1  \sigma_2-\dot{\sigma}_2  \sigma_1$.  
We know that the presence of a chemical potential in standard thermal equilibrium can lead to occupation of excited states with energy greater than the temperature, thereby overcoming Boltzmann suppression. Applying this intuition to the thermal-like nature of particle production in dS spacetime, we expect particles with masses $M \lesssim \lambda$ to be produced efficiently.\footnote{
More precisely, the inflationary couplings may contribute to the mass during inflation, in which case the condition for efficient production is effective mass, $\meff \lesssim \lambda$.}
Let us estimate the value of chemical potential in our case. From \cref{eq:Lambda_cutoff}, we see that $\lambda \simeq \dphiv/\Lambda \lesssim \dphiv^{1/2} \approx 60 H$, which is $\gg T_{\text{Hawking}}$. That is, we can expect that particles much heavier than $H$ can be produced without the dS analog of Boltzmann suppression, greatly widening the window
of  observable masses in the cosmological collider physics program.

In this paper, we demonstrate that the above expectations of 
non-Boltzmann-suppressed NG mediated by heavy scalars $H \ll M \lesssim \lambda $ can indeed be realized. The simplest way to see how this works is to first note that the ``chemical potential'' coupling can be thought of as $\chi$ being charged under a ``gauge'' field which is pure gauge in form, $A_{\mu} \equiv  \partial_{\mu} \phi/\Lambda $. It can therefore be removed by doing a gauge transformation on $\chi$. However, this will induce gauge transform phase factors 
in any other gauge-non-invariant masses or couplings in the theory:
\begin{align}\label{eq:fiedlredef}
    \mathcal{L} \supset \mathcal{O}(\chi, \chi^{\dagger}) + \text{c.c.}\rightarrow \mathcal{O}(\chi, \chi^{\dagger}) \, e^{i q \phi/\Lambda} + \text{c.c.} ,
\end{align}
where $\mathcal{O}$ is an interaction/mass in $\chi$ violating the fake ``gauge invariance'' by $q$ units (that is, there are $q$ more powers of $\chi$ than $\chi^{\dagger}$).
Given the slow-roll approximation, $\phi_0 \approx \dphiv t $, we see that we get effective high-frequency couplings of the heavy fields.
\begin{align}\label{eq:u1breaking}
    \mathcal{L} \supset \mathcal{O}(\chi, \chi^{\dagger}) \, e^{i q \lambda t} + \text{c.c.},
\end{align}
For example, if we depart from our assumption of equal heavy masses, or from $\lambda_2 =  \lambda_1$, there are effectively $\chi \chi$-type symmetry-breaking mass corrections accompanying the chemical potential, which will be multiplied by $e^{ 2i \lambda t}$ after the field redefinition. In general, in this way we get time-dependent $\chi$ couplings and mass terms with frequencies of order $\lambda$. These couplings can then naturally result in efficient particle production for masses up to $\lambda \gg H$.\footnote{It may be surprising to the reader that the chemical potential only overcomes dS Boltzmann suppression when the related charge is {\it not} conserved. However, in ordinary thermal equilibrium there is an analogous feature, in that it is crucial that the subsystem under study can gain/lose charge to the external thermal bath. If the subsystem charge is exactly conserved, then here too the chemical potential would have no effect.}
We will study the case of $\mathcal{O}(\chi,\chi^{\dagger})$ with $q=1$
and show that it can lead to striking NG signatures of the heavy scalars, the largest and the most theoretically tractable occurring via tree-level exchanges with the inflaton. These can be readily observable in upcoming LSS experiments \cite{Alvarez:2014vva, MoradinezhadDizgah:2017szk,MoradinezhadDizgah:2018ssw,Kogai:2020vzz} and future 21-cm cosmology \cite{Loeb:2003ya, Meerburg:2016zdz}. We will also briefly study the case $q=2$, which is mostly closely analogous to the fermionic case. In general, both $q=1$ and $q=2$ can co-exist, but for simplicity, we pursue them separately.

There are some qualitative differences between the heavy scalar case here and the earlier studies of heavy fields with non-zero spin $s$.
In these previous studies \cite{Garretson:1992vt, Barnaby:2010vf, Wang:2019gbi, Wang:2020ioa, Adshead:2015kza, Chen:2018xck, Hook:2019zxa, Wang:2019gbi, Hook:2019vcn}, the chemical potential modifies
the dispersion relation with a term linear in momentum using a coupling $\propto \vec{k}\cdot \vec{s}$. 
The modified frequency $\omega(t)$ violates the adiabaticity condition, i.e. $\dot{\omega}/\omega^2 \gg 1$, at a certain time during the evolution of mode functions, leading to particle production.
Of course a similar effect is impossible for spin-0 fields.
While these spin dependent effects are absent for scalar fields, a chemical potential can still manifest as an explicit time-dependence in mass corrections and couplings as seen in \cref{eq:u1breaking}, which is the source of particle production. Ref.~\cite{Wang:2019gbi} surveyed different chemical potential possibilities, but concluded that there was no effect for complex scalars as given by \cref{eq:twoScalar_chemPot}. In this paper, we will show that there is indeed a strong effect within theoretical control.

Our main focus will be on $q=1$ case with $\mathcal{O} = \chi$, which has the advantage that the contribution of the inflaton-scalar coupling to the bispectrum occurs at \emph{tree-level}.
This removes the loop-suppression seen in gauge boson and fermion cases, gives a robustly large mass reach up to $\dphiv^{1/2} \approx 60H$,
while allowing an analytically tractable calculation. 
Note this is significantly higher than the chemical potential reach for fermions $\simeq \dphiv^{1/4} H^{1/2} \approx 10H$.
While the spin-1 chemical potential reach is as high as the scalar case $M \simeq \dphiv^{1/2} \approx 60H$, getting a viable signal
requires a period of exponential growth in particle production \cite{Garretson:1992vt}, in turn needing a very close coincidence of the mass and chemical potential scales in order to stop in time.
By contrast in the scalar case, no exponential growth or coincidences are necessary.  
The meaning of mass is obscured in the presence of large chemical potential due to its breaking of Lorentz invariance. But we will carefully identify the physical mass after the field rotation which removes the chemical potential, as in \cref{eq:fiedlredef}, and show that strong signals can be obtained for physical masses $ \gg H$ without any fine-tuning of parameters.


The field redefinition leading to the form of Lagrangian, \cref{eq:fiedlredef}, demonstrates that the scalar chemical potential is equivalent to a special form of particle production due to non-derivative couplings to the inflaton, similar in spirit to \cite{Flauger:2016idt}. However there are some significant differences. Generally, non-derivative couplings will induce radiative corrections to the inflationary effective potential upon integrating out the heavy matter, of concern given the fragility of slow-roll inflation. Furthermore, non-derivative couplings of (nearly) massless scalars ordinarily yield divergent late-time correlators.
However, in our case the 
non-derivative couplings to the inflaton do not break the shift symmetry under which $\phi \rightarrow \phi + c$ when it is accompanied by the phase rotation $\chi \rightarrow e^{-ic/\Lambda} \chi$. This means that at the level of the inflationary effective potential, radiative corrections from integrating out matter must respect $\phi$ shift symmetry, and hence vanish. This matches the obvious conclusion from the original form of interaction, \cref{eq:twoScalar_chemPot}, where the derivative couplings of $\phi$ cannot generate an effective $\phi$ potential. Similarly, the manifest derivative couplings in \cref{eq:twoScalar_chemPot} ensure that all late-time divergences cancel when the correlators are computed using \cref{eq:fiedlredef} (which however is more useful in computing the effects of the chemical potential).
Finally, it may appear worrying that the non-derivative couplings in \cref{eq:fiedlredef} can be Taylor expanded in powers of $\phi/\Lambda$ even though the $\phi$ field transit is much greater than $\Lambda$. In general, non-derivative couplings containing $(\phi/\Lambda)^n$ would not have a controlled expansion. But here the symmetry requirement that the inflaton factor non-linearly carries ``charge'' $q$ uniquely gives the re-summed form $e^{i q \phi/\Lambda}$, thus matching the manifest EFT control in \cref{eq:EFT_operators}.

As seen in \cref{eq:u1breaking}, the particle production resulting from the chemical potential is continuous and approximately constant during inflation, given by a steady frequency of the effective couplings/masses, as opposed to a more time-dependent or punctuated particle production. Relatedly, the shift symmetry gives us an approximate time translation invariance, which combined with dS isometries implies that approximate scale invariance of our correlators is maintained. This structure of correlators is useful in doing the spectroscopy of the heavy particles from the bispectrum in a simple way.

The rest of the paper is organized as follows: we describe the relevant observables and fix the notation in \cref{Prelim}. Then in \cref{minModel}, we present our model that illustrates the mechanism of chemical potential in the context of heavy complex scalars. Section \ref{sec:bispectrum_approx} contains the main results of this paper where we calculate the bispectrum in the squeezed limit and analyze the strength of NG in various regions of the parameter space. There we also give a simple and intuitive estimate of the bispectrum using the method of stationary phase while a full calculation is carried out in \cref{bispectrum_full} using the results from \cref{sec:HankelInt}.
After discussing in \cref{Sec:Meff_extract}  
a novel procedure to infer the effective mass of the heavy particle during inflation from the momentum dependence of the bispectrum, we conclude in \cref{sec:conclusions}. 

\section{Preliminaries}\label{Prelim}
Let us start with the definitions and the notation that we will use throughout the paper. We will work in $(-,+,+,+)$ signature of the metric. Under the slow-roll approximation, the fractional change in the Hubble constant $|\dot{H}/H^2|$ is $\lesssim 1\%$ \cite{Akrami:2018odb}. Since $H$ changes slower than other quantities during inflation, we take it to be a constant and approximate the metric during inflation by the dS metric, 
\begin{align}
ds^{2}_{\text{dS}}
&= -dt^2 + a(t)^2 d\Vec{x}^2\\[0.3em]
&= \frac{-d\eta^2+d \vec{ x}^2}{(\eta H)^2},  
\end{align}
where $\eta$ is the conformal time. It is related to the proper time $t$ through the relation $ d\eta = dt/a(t) $, where $a(t)$ is the scale factor. This gives $\eta H = -e^{-Ht}$ during inflation. We set $H=1$ in the rest of the text for the ease of calculation, unless it is explicitly written. Factors of $H$ can be restored by dimensional analysis.\par
The inflaton field can be separated into a classical homogeneous background $\phi_0$ that sources inflation, and the (quantum) fluctuations $\xi$ that seed density perturbations, $\phi = \phi_{0}(t) + \xi(t,\vec{x})$. In the spatially flat gauge \cite{Maldacena:2002vr}, the fluctuations $\xi(t, \vec{x})$ are related to the gauge-invariant comoving curvature perturbation $\mathcal{R}$ (for a review and original references see e.g. \cite{Malik:2008im}) as
\begin{align}\label{R}
    \mathcal{R} = +\frac{\xi}{\dphiv} \;.
\end{align}
The observations require the primordial power spectrum $P_{\mathcal{R}}(k)$ to be (almost) scale invariant,
\begin{align}\label{PowerSpectrum}
    P_{\mathcal{R}}(k) \equiv \<\mathcal{R}_{\vec{k}} \mathcal{R}_{-\vec{k}} \>^\prime = \frac{1}{\dphiv^2}\<\xi_{\vec{k}} \xi_{-\vec{k}} \>^\prime =  \frac{1}{\dphiv^2}\frac{(k/k_{*})^{n_{s}-1}}{2k^3} \;,
\end{align}
where $n_{s} \approx 0.96$, $k_{*} = 0.05$ Mpc$^{-1}$ is the pivot scale, and  $\dphiv^{1/2} \approx 60 H$ \cite{Akrami:2018odb}. Here and below, the $'$ denotes the convention that momentum conserving factors have been taken out,
\begin{align}
    \<\mathcal{R}_{\vec{k}_1} \cdots \mathcal{R}_{\vec{k}_n}\> \equiv (2\pi)^3 \delta^{3}(\vec{k}_1+\cdots+\vec{k}_n)\,\<\mathcal{R}_{\vec{k}_1} \cdots \mathcal{R}_{\vec{k}_n}\>' \;.
\end{align}

The primordial non-gaussianity (NG) corresponds to having connected non-zero $n-$point correlation functions of $\mathcal{R}$. More specifically, the bispectrum is the 3-point function in momentum space denoted by $B(k_1,k_2,k_3)$, 
\begin{align}
    B(k_1,k_2,k_3) = \<\mathcal{R}_{\vec{k}_1} \mathcal{R}_{\vec{k}_2} \mathcal{R}_{\vec{k}_3}\>' \;.
\end{align}
It is a convention to denote the bispectrum by $F(k_1,k_2,k_3)$, which can be thought of as the bispectrum appropriately normalized by the power spectrum
\begin{align}\label{eq.Fdef}
    F(k_1,k_2,k_3) = \frac{5}{6}\frac{B(k_1,k_2,k_3)}{ \left( P_{\mathcal{R}}(k_1)\,P_{\mathcal{R}}(k_3) +P_{\mathcal{R}}(k_1)\,P_{\mathcal{R}}(k_2)+P_{\mathcal{R}}(k_2)\,P_{\mathcal{R}}(k_3) \right)} \;.
\end{align}
In our analysis, we will be mostly interested in the squeezed limit, i.e., when $k_1 \approx k_2\gg k_3$. Using \cref{R} and \cref{PowerSpectrum}, the expression for $F$ can be simplified in this limit as
\begin{align}\label{eq:Fsqueezed_def}
    F_{\text{squeezed}} \approx \frac{5}{12} \dphiv \frac{\< \xi_{\vec{k}_2}\xi_{\vec{k}_3}\xi_{\vec{k}_1} \>^\prime}{\< \xi_{\vec{k}_1} \xi_{-\vec{k}_1}\>^\prime\< \xi_{\vec{k}_3} \xi_{-\vec{k}_3}\>^\prime}.
\end{align}
It is customary to typify the strength of NG by a single number, $f_{\text{NL}}$, at the ``equilateral'' configuration where all the momenta have the same magnitude,
\begin{align}\label{fnl_def}
    f_{\text{NL}} \equiv F(k,k,k) = \frac{5}{18} \frac{B(k,k,k)}{P_{\mathcal{R}}(k)^2} \;.
\end{align}
This definition is consistent with the convention for local NG ($f_{\text{NL}}^{\text{\tiny local}}$) at the equilateral point. Using \cref{R} and \cref{PowerSpectrum}, this can be written in terms of inflaton correlation functions,
\begin{align}\label{fnl_simplified}
f_{\text{NL}} = \frac{5 }{18}\, \dphiv\, \frac{\< \xi_{\vec{k}_1}\xi_{\vec{k}_2}\xi_{\vec{k}_3} \>}{\< \xi_{\vec{k}} \xi_{-\vec{k}}\>^2} \;,
\end{align}
where $|\vec{k}_{1,2,3}|=|\vec{k}|$.

The constraints on $f_{\text{NL}}$ vary depending on the shape of the bispectrum. The latest Planck analysis sets the bounds as 
$|f_{\rm NL}|\sim \mathcal{O}(5-50)$ depending on the shape of NG
\cite{Akrami:2019izv}.
In the near future, LSS observations (see e.g.~\cite{Alvarez:2014vva}) are expected to probe $\delta f_{\text{NL}} \sim \mathcal{O}(1)$ --- bringing several of our benchmark scenarios under the LSS reach, as we will see later. 
More futuristic cosmic variance limited 21-cm cosmology experiments can reach even better sensitivities with $\delta f_{\rm NL} \sim \mathcal{O}(10^{-2})$ or even smaller \cite{Loeb:2003ya,Munoz:2015eqa}.

We use the in-in formalism to evaluate the bispectrum. The reader may refer to \cite{in-in_weinberg} for a more comprehensive review. Using this formalism, the expectation value of a Heisenberg operator $\hat{\mathcal{O}^H}$ at any single time $t$ can be perturbatively computed using,
\begin{align}\label{in-in}
    \langle \Omega |\, \hat{\mathcal{O}}^H (t) \,| \Omega \rangle =  \langle 0| \left(\bar{T}e^{i\int_{-\infty(1+i\epsilon)}^{t} H^{I}_{\text{int}} \,dt } \right) \hat{\mathcal{O}}^{I}(t) \left(T e^{-i\int_{-\infty(1-i\epsilon)}^{t} H^{I}_{\text{int}}\,dt }\right)|0 \rangle \;,
\end{align}
where we can expand in powers of the interaction Hamiltonian $H^{I}_{\text{int}}$ to any desired order. Here we see the appearance of both time- and anti-time ordered exponentials, respectively denoted by $T$ and $\Bar{T}$ operators. The state $|0\rangle$ is the free Bunch-Davies vacuum but the $i\epsilon$ prescription projects it onto the interacting vacuum $|\Omega \rangle$. The superscript $I$ denotes fields in the interaction picture. 
The bispectrum is the 3-point correlation function evaluated at the end of inflation, i.e., $\hat{\mathcal{O}}(t) \rightarrow (\xi_{\vec{k}_1}\xi_{\vec{k}_2}\xi_{\vec{k}_3})|_{\eta \approx 0}$ in \cref{in-in}.

\section{A minimal model}\label{minModel}
\subsection{Chemical potential for a complex scalar with U(1) symmetry breaking}\label{subsec:chemPot_general}
Before getting into our actual models, let us consider a simple case of a complex scalar field $\chi$ charged under a global $U(1)$, where we introduce an explicit chemical potential by hand. The conserved current $J_{\mu} = -i(\chi\partial_{\mu} \chi^{\dagger}  - \chi^{\dagger}\partial_{\mu} \chi  )$ can be used to introduce a non-zero chemical potential, $\lambda$, by shifting the Hamiltonian density as $\mathcal{H} \rightarrow \mathcal{H}-\lambda J_{0}$. 
Integrating out the canonical momentum in the partition function, $\mathcal{Z} \sim \int D \chi\, D \Pi \, \exp[-\int d^4 x \, (\mathcal{H}-\lambda J_{0})]\,$, gives the corresponding Lagrangian, $\mathcal{Z} \sim \int D \chi\, \exp[-\int d^4 x \,  \mathcal{L}_{\chi}]\, $ \cite{thermal_FT},
\begin{align}\label{ToyChemPot}
    \mathcal{L}_{\chi} \, &\supset \, -|\partial \chi|^2 - M^2 |\chi|^2 +i\lambda (\dot{\chi}^{\dagger} \chi -\dot{\chi} \chi^{\dagger}) + \lambda^2 |\chi|^2 \nonumber\\[0.3em]
    &= \, \left[ (\partial_{t}+i\lambda)\chi \right] [ (\partial_{t}-i\lambda)\chi^{\dagger}] -|\partial_{i}\chi|^2- M^2|\chi|^2 \:.
\end{align}
Note that adding a chemical potential in the Hamiltonian is equivalent to introducing coupling of the complex field to an external gauge potential $A_{\mu}$ with a non-zero time component $A_{0}$.
Then $A_{0} (\equiv \lambda)$ is the chemical potential.
It is important to clearly identify the physical mass in terms of the various mass parameters in the Lagrangian, since it is obscured by the Lorentz-violating nature of the chemical potential. We will address this question shortly.

In the standard thermodynamic setting, the subsystem under consideration can gain or lose charges to the heat bath, which is key to chemical potential having physical effect. In the inflationary setting on the other hand, the subsystem is the entirety of space and hence, the charge conservation is exact. 
This can be easily seen by noting that any non-trivial effect of the chemical potential can be eliminated by a suitable field rotation, $\chi \rightarrow e^{-i\lambda t} \tchi$.\footnote{When the subsystem is coupled to a thermal bath, the non-zero time-component of the gauge field can not be rotated away due to periodic boundary conditions on the fields in the subsystem.}
Nevertheless, non-conservation due to the heat bath can be replaced here by small explicit breaking of the $U(1)$ symmetry.
Consider the simplest case of adding a linear symmetry-breaking term as follows,
\begin{align}
    \mathcal{L}_{\chi} \supset  \left[ (\partial_{t}+i\lambda)\chi \right] [ (\partial_{t}-i\lambda)\chi^{\dagger}] -|\partial_{i}\chi|^2 - M^2|\chi|^2 - \alpha\, (\chi+\chi^{\dagger}) \;.
\end{align}
After the field rotation as before $\chi \rightarrow e^{-i\lambda t} \tchi$, we see that the symmetry-breaking term retains the effect of chemical potential in the form of a time-dependent phase,
\begin{align}\label{eq:ComplexScalar_rotated}
    \mathcal{L}_{\tchi} \supset -|\partial \tchi|^2  - M^2|\tchi|^2 - \alpha\, (\tchi e^{-i\lambda t}+\tchi^{\dagger}e^{+i\lambda t}) \;.
\end{align}
For two independent real scalar fields ($\sigma_1,\, \sigma_2$) as in \cref{eq:twoScalar_chemPot}, such symmetry-breaking terms are naturally present when written in terms of $\chi = \sigma_1 + i \sigma_2$.

In our full model, we will realize such a residual time-dependence with frequency $\lambda$, coupled to inflaton fluctuations,  capable of producing heavy particles in association with inflatons at energies of order $\lambda$. But we can already identify the physical mass, having eliminated the chemical potential in favor of the time-dependent $\alpha$ coupling.  Treating the linear term in \cref{eq:ComplexScalar_rotated} as a perturbation for small $\alpha$, we see that the physical mass is nothing but $M$ (e.g., $M^2$ term is the standard dS-invariant if we are expanding around dS-spacetime). We will generalize this approach to identifying the physical mass after field re-definition and computing the NG signals as a function of it in what follows.


\subsection{Realizing a chemical potential via inflaton couplings}
Coming to the case of inflation, the chemical potential can be implemented with derivative coupling of inflaton with the complex scalar current $J^\mu$. Consider the Lagrangian,
\begin{align}\label{L}
    \mathcal{L} = \sqrt{-g} \left\lbrace \vphantom{\frac{1}{2}} \right.& \left. -\frac{1}{2}(\partial \phi)^2 - V(\phi) 
    -|\partial \chi|^2 
		-M^2 |\chi|^2  \right. \nonumber\\[0.3em]
	 & \left. 
	 -\frac{i \partial_{\mu} \phi}{\Lambda} \left(\chi \partial^{\mu} \chi^{\dagger}\, -  \chi^{\dagger}\partial^{\mu} \chi\, \right)
	-  \frac{c\,(\partial \phi)^2 }{\Lambda^{2}} |\chi|^2
	+\alpha \left(\chi+\chi^\dagger\right) \, \right\rbrace \;.
\end{align}
Here $V(\phi)$ is some suitable inflationary potential whose details we do not need to specify.
The second line describes the interactions, out of which the first term is the coupling of the inflaton to the current. In the slow-roll approximation, the inflaton gets a VEV such that $\dphiv \simeq (60H)^2$ is a constant to a good approximation. This generates a chemical potential, 
\begin{align}
    -\frac{1}{\Lambda}\partial_{\mu}\phi J^{\mu} \supset  \frac{\dphiv}{\Lambda}J_{0} = \lambda J_{0},
\end{align}
where $\lambda \equiv \dot{\phi}_{0}/\Lambda$. 
It is evident from the discussion above that we must also include a small explicit symmetry-breaking term, which is the last term on the second line in \cref{L}.
The second term, with the coefficient $c$, should be naturally present as it is consistent with the shift symmetry of inflaton as well as the $U(1)$ symmetry of the complex scalar. 
The coefficient of this term gets modified as it is also generated from the kinetic term for $\chi$ after the field redefinition that removes the dim-5 inflaton-current coupling,
\begin{align}\label{redef}
    \chi = e^{-i\phi/\Lambda} \, \tchi \;.
\end{align}
Then \cref{L} can be written as
\begin{align}\label{Lrotated}
    \mathcal{L} = \sqrt{-g} \left\lbrace   
    \vphantom{\frac{1}{2}} \right.& \left. -\frac{1}{2}(\partial \phi)^2 - V(\phi) 
    -|\partial \tchi|^2 -M^2 |\tchi|^2
     \right. \nonumber\\[0.3em]
	 & \left. 
	+ (1-c)\, \frac{(\partial \phi)^2 }{\Lambda^{2}} |\tchi|^2
	+\alpha \left(\tchi e^{-i\phi/\Lambda}+\tchi^\dagger e^{+i\phi/\Lambda}\right) \right\rbrace  \;.
\end{align}
In the new field variable $\tchi$, the chemical potential gets translated into a non-trivial phase in the symmetry-breaking term. Another advantage of this step is that the Lagrangian in \cref{Lrotated} no longer has interactions with time-derivatives $ \partial_{0} \tchi$. 
It is then straightforward to get the interaction Hamiltonian starting from the Lagrangian density using $\mathcal{H}_{\text{int}}= -\mathcal{L}_{\text{int}}$.\footnote{It can be shown that the additional terms with $(\partial \phi)^2$ give rise to extra interaction terms only at fourth-order in perturbation or higher \cite{Wang13:LtoH}. Since we are only looking at diagrams at tree level, this will not affect our analysis.} For simplicity and perturbative control, we restrict the coefficient $|c| \leq 1$. Note that $c=1$ is a case that corresponds to the addition of chemical potential without affecting the mass,
as seen in \cref{ToyChemPot}. Here, we keep this coefficient arbitrary. 

\subsection{Analysis of the spatially homogeneous VEV of the complex scalar} \label{homSol}
In the following, we will restrict our attention to the parameter space where interactions in \cref{Lrotated} are perturbative such that the inflaton dynamics is not altered at leading order in perturbation theory. Therefore we can treat the homogeneous part of the inflaton as $\phi_0$ where $\dot{\phi}_0\approx (60H)^2$.
This perturbativity is ensured by taking $\Lambda$($\alpha$) sufficiently large(small), as discussed in more detail in \cref{Constraints}. The term linear in $\tilde{\chi}$ in \cref{Lrotated} breaks the $U(1)$ symmetry explicitly and gives a \textit{time-dependent} VEV $\tchi_0$ which we estimate now.  

From \cref{Lrotated}, the Lagrangian terms involving $\tchi$ are
\begin{equation}\label{LchiQuad}
\begin{aligned}
    \mathcal{L}_{\tchi} & \supset  \sqrt{-g} \left\lbrace -|\partial \tchi|^2- \left(\meff^2 + {\cal O}(\xi)\right)\, |\tchi|^2 +\underbrace{\alpha \left( \tchi \, e^{-i(\phi_{0}+ \xi)}/\Lambda + \rm{h.c.} \right)} \, \right\rbrace \\[0.3em]
    & \hspace{17em}\alpha \left( \tchi \, e^{-i\lambda t} \,(1+ {\cal O}(\xi)) + \rm{h.c.} \right)
\end{aligned}
\end{equation}
where we have expanded the inflaton field around its VEV with $\xi$ denoting the inflaton fluctuations. In the second line, we have used the fact that $\dphiv \approx {\rm constant}$ during inflation in the slow-roll approximation, and hence, we can write
\begin{align}
    \frac{\phi_0}{\Lambda} \approx \frac{\dphiv }{\Lambda}(t - t_i) = \lambda\, (t-t_i)
\end{align}
Here, $t_i$ depends on the initial value of $\phi_0$. Its exact value will not affect our final results as it only appears as an unimportant overall phase (hence, not shown in \cref{LchiQuad}). The effective mass of the heavy field can be written as 
\begin{align}
M^{2}_{\text{eff}} = M^2 + (1-c) \, \lambda^2.  
\end{align}
Following the discussion in \cref{subsec:chemPot_general}, the physical (approximately) dS-invariant mass is most readily identified in $\tchi$ variable, where the chemical potential term is redefined away. 
In the limit where $\alpha$ perturbation is switched off, $\meff$ is clearly the physical mass during inflation. 
For $\alpha =0$, the heavy particle production is Boltzmann suppressed by $e^{-\pi \meff / H}$ for $\meff \gg H$ in standard fashion, whereas it is produced (in association with inflatons) without any such suppression in the presence of non-zero $\alpha$ for $\lambda \gtrsim \meff > H$, due to the high $\lambda$-frequency time-dependence of the $\alpha$ coupling. In particular, no fine-tuned cancellation between various parameters in $\meff$ is needed to obtain an unsuppressed signal.

The parameter $c$ controls the contribution of chemical potential to the effective mass, as well as symmetry-preserving couplings of the heavy particle to the inflaton. As mentioned before, $c=1$ is a case where the coupling of inflaton to the current only generates a chemical potential and does not affect the physical mass, i.e., $\meff = M$.
For any value of $c$, we expect the range of values $-(1-c) \lambda^2 < M^2 < c \lambda^2$, corresponding to $ H< \meff < \lambda$, to give unsuppressed signal (we will see this explicitly later). Note that this range includes tachyonic $M^2<0$ for $c<1$. However, it does not lead to instability as long the physical mass $\meff^2>0$.
The meaning of tachyonic $M$ is only seen after inflation ends, and $\phi$ decays and lies at the minimum of its potential today, $\dot{\phi} \propto \lambda = 0$, so $M$ represents the post-inflation physical mass. Being tachyonic, $M^2 < 0$, indicates only that $\chi$ will be unstable to acquiring a post-inflation VEV. But this does not affect our inflationary analysis of NG produced.

Due to the time-dependent phase in the term linear in $\tchi$,
the standard procedure of shifting the field by a constant value will not get rid of this tadpole completely. We instead solve the equation of motion (EOM) of $\tchi_{0}$ following from \cref{LchiQuad}. 
Let us write the high-frequency phase in $\alpha$-coupling using conformal time as
\begin{align}
   e^{\pm i \lambda (t-t_i)} = e^{\pm i \lambda \log(\eta/\eta_i)} \;,
\end{align}
From \cref{LchiQuad}, the EOM for the complex scalar field (in the dS limit of inflationary spacetime) is 
\begin{align}\label{complex_eom}
   \left[ \partial_{\eta}^{2} \,  - \frac{2}{\eta} \, \partial_{\eta} + \left( -\nabla^2 + \frac{M^{2}_{\text{eff}}}{\eta^2} \right) \right] \tchi(\vec{x},\eta) \,=\,  +\alpha \,  \frac{(\eta/\eta_{i})^{-i\,\lambda}}{\eta^2} \;.
\end{align}
This is a standard EOM for a massive scalar field in the inflationary background with an extra time-dependent piece on the RHS coming from the linear term. This extra piece will source spatially homogeneous part, i.e., $\vec{k}=0$ mode. We write $\tchi \rightarrow \tchi_{0}(\eta) + \deltchi(\vec{x},\eta)$, where $\tchi_{0}$ is the time-dependent VEV, and $\deltchi(\vec{x},\eta)$ is the fluctuation. Putting $\nabla^2 \tchi_{0} =0 $ in \cref{complex_eom}, we get the equation for the homogeneous part,
\begin{align}
   \left[ \partial_{\eta}^{2} \,  - \frac{2}{\eta} \, \partial_{\eta} +\frac{M^{2}_{\text{eff}}}{\eta^2}  \right] \tchi_{0}(\eta) \,=\,  +\alpha \,  \frac{(\eta/\eta_{i})^{-i\,\lambda}}{\eta^2} \;.
\end{align}
The general solution can be written as,
\begin{align}
    \tchi_{0}(\eta) = C_{\pm}\, (-\eta)^{3/2 \,\pm \,i\, \mu} \, + \, \alpha \,\frac{(\eta/\eta_i)^{-i\,\lambda}}{(M^{2}_{\text{eff}}-\lambda^2+3i\lambda)} \;,
\end{align}
where $\mu = \sqrt{M^{2}_{\text{eff}} - 9/4}$. The first term corresponds to oscillations of the massive field given by its mass $\sim \mu$, and the usual $1/\sqrt{\rm volume}$ dilution of mode functions. Due to the dilution, this term quickly becomes negligible. On the other hand, the second term exhibits a forced-oscillator type behaviour coming from the $\tchi$-tadpole in \cref{LchiQuad}, which gives time-dependence to the VEV $\tchi_{0}$. This term does not dilute (nor diverge) in late time and vanishes at the end of inflation as $\dot{\phi}_{0} \rightarrow 0$. Thus, the homogeneous part of the complex field can be written as
\begin{align}\label{chi0}
    \tchi_{0}(\eta) \,\approx\, \frac{\alpha \, (-\eta)^{-i\,\lambda}\, e^{i\theta}}{\sqrt{(M^{2}_{\text{eff}}-\lambda^2)^2+9\lambda^2}} \equiv \kappa^\dagger (-\eta)^{-i\,\lambda} \;,
\end{align}
where $\theta = \tan^{-1}(-3\lambda/(\meff^2-\lambda^2)+\lambda\ln(-\eta_i))$ is a constant phase. 
Notice that the amplitude of the oscillating VEV in \cref{chi0} depends on the detuning between the natural frequency $\meff$ and the external frequency $\lambda$, just like the case of a forced oscillator. The damping comes from the Hubble friction $\sim \frac{1}{\eta}\partial_\eta\tchi$ in \cref{complex_eom}. 

An important consequence of the time-dependent VEV for the heavy complex scalar is that the leading NG will now be at \emph{tree-level}, which means we no longer pay the $\sim 1/(16\pi^2)$ loop suppression, and the calculation is a lot more tractable as compared to the loop-level calculations in scenarios involving chemical potential for fermions and gauge bosons.
\subsection{Quantization and mode functions of the complex scalar}\label{modeFunc}
After doing the analysis for the homogeneous solution, let us move on to discuss the fluctuations. Taking a Fourier transform in \cref{complex_eom}, we get the equation for the $\vec{k}$-th mode ($\vec{k}\neq 0$) as  
\begin{align}\label{EOM_chi_k}
    \left[ \partial_{\eta}^{2} \, - \frac{2}{\eta} \, \partial_{\eta} + \left( k^2 + \frac{M^{2}_{\text{eff}}}{\eta^2} \right) \right] \deltchi(\vec{k},\eta) \,=\, 0 \;.
\end{align}
Here, we have ignored a quadratic mixing between $\deltchi$ and $\xi$. We will later take this into account perturbatively in the insertion approximation.
The RHS in \cref{complex_eom}
is purely spatially homogeneous and does not affect $\Vec{k}\neq 0$.
The equation for $\deltchi(\vec{k},\eta)$ is the standard EOM for a massive scalar field in the dS limit of inflationary spacetime. In terms of the redefined field $\tchi$, the effect of the chemical potential is solely on the time-dependent VEV $\tchi_{0}$, while the fluctuations follow the same dynamics as a free massive scalar field. The general solution is
\begin{align}\label{genM_sol}
    \deltchi(\Vec{k},\eta) = c_{1}\, (-\eta)^{3/2} H^{(1)}_{i\mu}(-k\eta) \,+\, c_{2}\, (-\eta)^{3/2} H^{(2)}_{i\mu}(-k\eta) \;.
\end{align}
Here, $H^{(1\text{ or }2)}_{i\mu}$ are the Hankel functions. The quantized field can be taken to have the following form,
\begin{align}\label{chitsol}
    \deltchi(\Vec{k},\eta) = \bgk{k} a_{-\Vec{k}} + f_{k}(\eta)\, b_{\Vec{k}}^{\dagger} \;,
\end{align}
where $a_{\Vec{k}} \, (a^{\dagger}_{\Vec{k}})$ and $b_{\Vec{k}} \, (b^{\dagger}_{\Vec{k}})$ are the destruction (creation) operators for particle and anti-particle excitations, respectively, i.e. $a_{\vec{k}}|0\> = b_{\vec{k}}|0\> =0$, where $|0\>$ is the Bunch-Davies (BD) vacuum. 
Imposing canonical quantization, using the conjugate momentum $\pi(\vec{x},\eta)$, $[\deltchi(\Vec{x},\eta) , \pi(\Vec{x},\eta)] = +i\, \delta^{3}(\Vec{x}-\Vec{y})$ translates into standard commutation relations $[a_{\Vec{k}},a^{\dagger}_{\Vec{k'}}] = (2\pi)^3 \delta^{3}(\Vec{k}-\Vec{k'})$ and $[b_{\Vec{k}},b^{\dagger}_{\Vec{k'}}] = (2\pi)^3 \delta^{3}(\Vec{k}-\Vec{k'})$, given that the mode functions satisfy the Wronskian condition 
\begin{equation}
\bar{g}_k g_k^\prime-f_k\bar{f}_k^\prime =  i\eta^2 \;,
\end{equation}
where $'$s denote derivatives with respect to conformal time $\eta$.

Treating all the interactions perturbatively, the quadratic part of the Lagrangian for fluctuations corresponds to a free complex scalar in dS space-time.
Thus, minimization of Hamiltonian requires choosing appropriate positive frequency mode at early time, i.e., $f_{k} , g_{k} \propto e^{i k \eta} $ at $\eta\rightarrow -\infty$. These initial condition and the Wronskian condition fix the form of the mode functions $g_k,f_k$ to be
\begin{equation}\label{chi_modefunc}
    \begin{aligned}
    \bar{g}_k &= N^{*}_{g}(-\eta)^{3/2}H_{i\mu}^{(1)}(-k\eta)\\[0.3em]
    f_k &= N_f(-\eta)^{3/2}H_{i\mu}^{(2)}(-k\eta) \;,
    \end{aligned}
\end{equation}
where $N_f=\frac{\sqrt{\pi}}{2}e^{\pi\mu/2}e^{-i\pi/4}$ and $N^{*}_{g}=\frac{\sqrt{\pi}}{2}e^{-\pi\mu/2}e^{+i\pi/4}$. It can be checked that $g_k=f_k$. Furthermore,
\begin{equation}
    \begin{aligned}
    \bar{g}_k(\eta\rightarrow -\infty)=\frac{1}{\sqrt{2k}}(-\eta)e^{-ik\eta}\\
f_k(\eta\rightarrow-\infty)=\frac{1}{\sqrt{2k}}(-\eta)e^{ik\eta},
    \end{aligned}
\end{equation}
as required by imposing BD initial conditions.

Assuming the validity of perturbative analysis (discussed in \cref{Constraints}), we can then take the standard $M\rightarrow 0$ limit for inflaton. Quantizing inflaton fluctuation as
\begin{align}
    \xi(\vec{k},\eta) = \bar{u}_{k} c_{-\vec{k}} + u_{k} c^{\dagger}_{\vec{k}} ,
\end{align}
we get the mode functions by setting $\mu = 3i/2$ in \cref{chi_modefunc},
\begin{equation}\label{infl_modefunc}
    \begin{aligned}
    u_{k}(\eta) &= \frac{(1-i k \eta)\, e^{i k \eta}}{\sqrt{2\, k^{3}}}\\
    \bar{u}_{k}(\eta) &= \frac{(1+i k \eta)\, e^{-i k \eta}}{\sqrt{2\, k^{3}}} \;.
\end{aligned}
\end{equation}

\subsubsection{Boltzmann-like suppression for fields with no chemical potential}\label{subsubsec:boltzmann_supp}
Having studied the mode functions, let us take a detour to understand the origin of the Boltzmann suppression in the absence of chemical potential. 
Consider a heavy real scalar field $\sigma$ with mass $M$ in dS space-time. The free field EOM is the same as in \cref{EOM_chi_k}. The scalar field is then quantized using BD initial condition as,
\begin{align}
    \sigma = \bar{f}_{k} a_{-\vec{k}} + f_{k} a_{\vec{k}}^{\dagger} \qquad \text{where} \quad \bar{f}_{k} = N_{f}^{*} (-\eta)^{3/2} H^{(1)}_{i\mu}(-k\eta) \,;  \quad \mu = \sqrt{M^2-9/4} \;. 
\end{align}
The BD initial condition means that at early time, the modes were deep into the horizon and hence look like Minkowski modes in the appropriate time variable. This is reflected in the early time limit of the mode functions, 
\begin{align}
    \bar{f}_k \xrightarrow{\eta \rightarrow -\infty} \bar{f}_{k}(in) = \frac{1}{\sqrt{2 k}} (-\eta)e^{-ik\eta},
\end{align}
where the \emph{positive} frequency solution is chosen as it minimizes the Hamiltonian at early times. In a time-dependent space-time, the same mode may not be the \emph{local} positive frequency mode at late times. This point becomes apparent in the late time limit where (up to some unimportant phases)
\begin{align}\label{eq:fout}
\bar{f}_k \xrightarrow{\eta \rightarrow 0} \bar{f}_{k}(out) &\simeq 
\frac{(-\eta)^{3/2}}{2\sqrt{\pi}}\left(\Gamma(-i\mu)(-k\eta/2)^{i\mu}e^{\pi \mu/2}+\Gamma(i\mu)(-k\eta/2)^{-i\mu}e^{-\pi \mu/2}\right) \nonumber\\[0.3em]
&\xrightarrow{M\gg H} \, \frac{(-\eta)^{3/2}}{\sqrt{2 \mu}}\left( (-k\eta/2)^{i\mu}+e^{-\pi \mu}(-k\eta/2)^{-i\mu}\right),
\end{align}
where in the second line, we have used 
\begin{align}
|\Gamma(i\mu)|\sim \sqrt{2\pi}|\mu|^{-1/2}e^{-\pi|\mu|/2}.
\end{align}
Notice that a small \emph{negative} frequency part, $(-\eta)^{-i \mu} \sim e^{i \mu t}$, has developed,
with a relative amplitude $e^{-\pi \mu}$, which corresponds to `cosmological particle production'. Then the probability of particle production is $\sim e^{-2\pi \mu}$, which is approximately $e^{-M/{T_{\rm Hawking}}}$, where the cosmological Hawking Temperature is given by $T_{\rm Hawking} = H/(2\pi)$. 
The non-analytic signature in the bispectrum comes from the `on-shell' production and propagation of heavy particles, which is always accompanied by this exponential suppression unless there is another energy source of particle production.
In our model, we create just such a source by harnessing the kinetic energy of the inflaton through its coupling to the complex current. This introduces an additional explicit time-dependence with frequency $\lambda \gg H$, and hence produces heavy fields more efficiently. 

\subsection{Interactions of the complex scalar with the inflaton}\label{interactions}
We focus on the interactions contributing to the bispectrum at tree-level, which is the dominant contribution in our model. These include terms that mix the complex scalar field with inflaton fluctuations, $\mathcal{H}_{\text{mix}}$, and 3-point interactions, $\mathcal{H}_{3}$. The relevant terms in the Lagrangian can be obtained from \cref{Lrotated} after expanding the phase of $\tchi$,
\begin{align}
    e^{i\,\phi/\Lambda} \,=\, e^{i\,(\phi_{0} + \xi)/\Lambda}  \,\approx\, e^{-i\lambda\, \tlog(\eta/\eta_i)} \left(1 + i\frac{\xi}{\Lambda}+...\right)\;,
\end{align}
and using the VEV of $\tchi$ from \cref{chi0}. The quadratic mixing terms are
\begin{align}\label{Hmix}
    \mathcal{H}_{\rm mix} = -\mathcal{L}_{\rm mix} = \beta_1\, (-\eta)^{i\lambda} \,\dot{\xi} \,\deltchi \,+\,\beta_2 \,(-\eta)^{i\lambda} \,\xi \,\deltchi+\text{h.c.} \;,
\end{align}  
where
\begin{align}\label{Hmix_coeff}
    \beta_1 = +\frac{2(1-c)\dphiv \kappa}{\Lambda^2},\hspace{2em} \beta_2 = i\frac{\alpha}{\Lambda}.
\end{align}
The relevant 3-point interaction terms are given by,
\begin{align}\label{Hint3}    
\mathcal{H}_{3} = -\mathcal{L}_{3} =
\rho_1 \, (-\eta)^{i\lambda}\, (\partial\xi)^2\, \deltchi+\rho_2 \, (-\eta)^{i\lambda}\,\xi^2\, \deltchi+\text{h.c.}\;,
\end{align}
where
\begin{align}\label{Hint3_coeff}
\rho_1=-\frac{(1-c)\kappa}{\Lambda^2},\hspace{2em} \rho_2=\frac{\alpha}{2\Lambda^2}\;.
\end{align}
Here we have read off the interaction terms directly from the Lagrangian using $\mathcal{H}_{\text{int}} = - \, \mathcal{L}_{\text{int}}$, which holds at third order in fluctuations (as explained 
after \cref{Lrotated}).
The extra time dependence $(-\eta)^{\pm i\lambda} $ in the vertices is the most important consequence of the chemical potential. It effectively injects (removes) energy of order $\lambda$ at each vertex, making the production of fields in the entire range $H < \meff \lesssim \lambda$ possible without the dS Boltzmann suppression.

\section{Approximate calculation of the bispectrum in the squeezed limit}\label{sec:bispectrum_approx}
Having worked out the interaction terms, we now evaluate the bispectrum following \cref{in-in},
\begin{align}\label{infl_in-in}
     \<\, \xi_{\vec{k_{1}}} \xi_{\vec{k_{2}}} \xi_{\vec{k_{3}}} \,\>'|_{\eta \rightarrow 0} =  \langle 0| \left(\bar{T}e^{i\int_{-\infty(1+i\epsilon)}^{0} H^{I}_{\text{int}} \,d\eta_1 } \right)  \xi_{0,\vec{k_{1}}} \xi_{0,\vec{k_{2}}}\xi_{0,\vec{k_{3}}} \left(T e^{-i\int_{-\infty(1-i\epsilon)}^{0} H^{I}_{\text{int}}\,d\eta_2 }\right)|0 \rangle \;,
\end{align}
where $\xi_{0,\vec{k}}$ is the inflaton fluctuation with comoving momentum $\vec{k}$ at $\eta\rightarrow 0$.
By expanding the exponential operator, it is easy to see that the bispectrum vanishes at zeroth order and the leading contribution comes from tree diagrams. 
While in general, the bispectrum contains both analytic and the non-analytic pieces, it is the latter that encode distinctive on-shell effects of the massive particle. Hence, for the purposes of this paper, we will compute only the non-analytic parts, keeping in mind that for a complete analysis one would also need to take into account the analytic contributions. However, the analytic contributions can be estimated to be comparable to our non-analytic estimates below in the regime of mild squeezing, i.e., analytic and non-analytic contributions to $\fnl$ are comparable.
The general form can be written as, 
\begin{align}\label{3pnt_schematic}
    &\<\, \xi_{\vec{k_{1}}} \xi_{\vec{k_{2}}} \xi_{\vec{k_{3}}} \,\>'|_{\eta \rightarrow 0} = I_{++}+I_{+-}+I_{-+}+I_{--}, \nonumber\\[0.3em]
    & I_{\pm\pm} \propto (\mp i) (\mp i)\int \frac{d\eta_1}{\eta_{1}^{4}} \int \frac{d\eta_2}{\eta_{2}^{4}} \,
    h(\eta_1,\eta_2)\,
    \tilde{d}_{\pm}(k_1,k_2;\eta_1)\, d_{\pm}(k_3;\eta_2)\,\times \nonumber\\ 
    & \hspace{13em} \lbrace G_{\pm \pm}(k_3; \eta_1,\eta_2) + F_{\pm \pm}(k_3; \eta_1,\eta_2)\rbrace\;.
\end{align}
The `$+$' indicates that the vertex that comes from $e^{-i\int H_{\text{int}} d\eta}$ with time ordering, while `$-$' comes from $e^{+i\int H_{\text{int}} d\eta}$ with anti-time ordering. Note that the first sign on $I_{\pm\pm}$ corresponds to the cubic interaction vertex, and the second sign corresponds to the inflaton-scalar mixing vertex. The function $h(\eta_1,\eta_2)$ captures the explicit time-dependence of the form $(-\eta)^{\pm i \lambda}$  in the vertex coefficients in \cref{Hmix,Hint3}. The functions $d_{\pm}$, $\tilde{d}_{\pm}$ are the inflaton bulk-boundary propagators, and they can be evaluated using \cref{infl_modefunc} as
\begin{equation}
\begin{aligned}
d_+(k_3;\eta_2) &\equiv \langle\xi(\eta_0)\xi(k_3,\eta_2)\rangle |_{\eta_0 \rightarrow 0} = \frac{1}{2k_3^3}(1-ik_3\eta_2)e^{ik_{3}\eta_2}\\
\tilde{d}_+(k_1,k_2;\eta_1) &\equiv \langle\xi(\eta_0)\xi(\eta_0)\xi(\eta_1,k_1)\xi(\eta_1,k_2)\rangle|_{\eta_0 \rightarrow 0} = \frac{1}{4k_1^3k_2^3}(1-ik_1\eta_1)(1-ik_2\eta_1)e^{ik_{12}\eta_1},  
\end{aligned}
\end{equation}
where $d_-$($\tilde{d}_-$) is the conjugate of $d_+$($\tilde{d}_+$), and $k_{12}=k_1+k_2$.
$G_{\pm\pm}$ and $F_{\pm\pm}$ are bulk-bulk propagators for the complex scalar field. These can be evaluated using \cref{chitsol} as
\begin{equation}\label{chi_propagator}
    \begin{aligned}
    G_{++}(k;\eta_1,\eta_2) &\,=\, \< \deltchi(\eta_1) \deltchi^{\dagger}(\eta_2)\>_{++} =
     \bar{f}_{k}(\eta_2) f_{k}(\eta_1) \theta(\eta_2-\eta_1) + \bar{g}_{k}(\eta_1) g_{k}(\eta_2) \theta(\eta_1-\eta_2),\\
    G_{+-}(k;\eta_1,\eta_2) &\,=\, \<\deltchi(\eta_2) \deltchi^{\dagger}(\eta_1)\>_{+-} =  
    \bar{g}_{k}(\eta_2) g_{k}(\eta_1),\\
    F_{++}(k;\eta_1,\eta_2) &\,=\, \< \deltchi^{\dagger}(\eta_1) \deltchi(\eta_2)\>_{++} =
    \bar{f}_{k}(\eta_1) f_{k}(\eta_2) \theta(\eta_1-\eta_2) + \bar{g}_{k}(\eta_2) g_{k}(\eta_1) \theta(\eta_2-\eta_1),\\
    F_{+-}(k;\eta_1,\eta_2) &\,=\, \<\deltchi^{\dagger}(\eta_2) \deltchi(\eta_1)\>_{+-} =  
    \bar{f}_{k}(\eta_2) f_{k}(\eta_1) \;.
\end{aligned}
\end{equation}
The remaining propagators can be related to these using relations,
\begin{align}
    G_{--} = (G_{++})^{*} \; ,\quad G_{-+} = (G_{+-})^{*} \; ,\quad F_{--} = (F_{++})^{*} \; ,\quad F_{-+} = (F_{+-})^{*}.
\end{align}
We will fix the proportionality factors in the above by using the relevant coupling parameters in a moment.
$I_{-+}$ and $I_{--}$ are complex conjugates of $I_{+-}$ and $I_{++}$, respectively. Thus, we only evaluate $I_{+-}$ and $I_{++} $ explicitly.

We compute the full analytic form of these contributions in \cref{bispectrum_full}, however next we show the same calculation using the stationary phase method, which is more transparent in terms of demonstrating particle production through chemical potential. It also proves to be a good approximation in the parameter regime of our interest, $\lambda \gg \meff > H$, as we will see in \Fig{stationary_phase}.

\subsection{Dominant contribution: $I_{+-}$ diagrams}\label{Ipm_stat}
Let us start with the evaluation of $I_{+-}$ diagram. 
For the purpose of illustrating the mechanism, we focus on the sub-diagram consisting of vertices with larger coefficients, i.e., $\beta_2$ and $\rho_2$.
We will see later that the NG estimate thus obtained matches the full analytic calculation very well. 

Using the form of vertices in \cref{Hmix,Hint3}, and \cref{3pnt_schematic} the sub-diagram has the form
\begin{align}\label{eq:Ipm_expression}
    I_{+-} &\supset (-i) (+i)\int \frac{d\eta_1}{\eta_{1}^{4}} \int \frac{d\eta_2}{\eta_{2}^{4}} \,   \tilde{d}_{+}(k_1,k_2;\eta_1) d_{-}(k_3;\eta_2)\times\, \nonumber\\[0.3em]
    &\hspace{6em} 
    \left\lbrace \beta_2 \rho^{*}_{2}\, \left(\frac{\eta_2}{\eta_1}\right)^{i\lambda}G_{+-}(k_3; \eta_1,\eta_2) + \beta^{*}_{2} \rho_2\, \left(\frac{\eta_2}{\eta_1}\right)^{-i\lambda}F_{+-}(k_3; \eta_1,\eta_2) \right\rbrace\;.
\end{align}
Looking at \cref{chi_propagator}, we see that the first term on the second line corresponds to the contraction $\<\deltchi \deltchi^{\dagger}\>$, describing particle propagation, while the second corresponds to $\<\deltchi^{\dagger} \deltchi\>$, describing anti-particle propagation. 
Without loss of generality, let is take $\lambda>0$ in the following analysis. We will see that with this choice, the contribution from the term with $F_{+-}$ propagator dominates. It will become clear that if $\lambda<0$, the analysis for $F_{+-}$ and $G_{+-}$ is simply switched, but our conclusions remain unchanged.

\begin{figure}[t]
	\centering
		\includegraphics[width=0.9\textwidth]{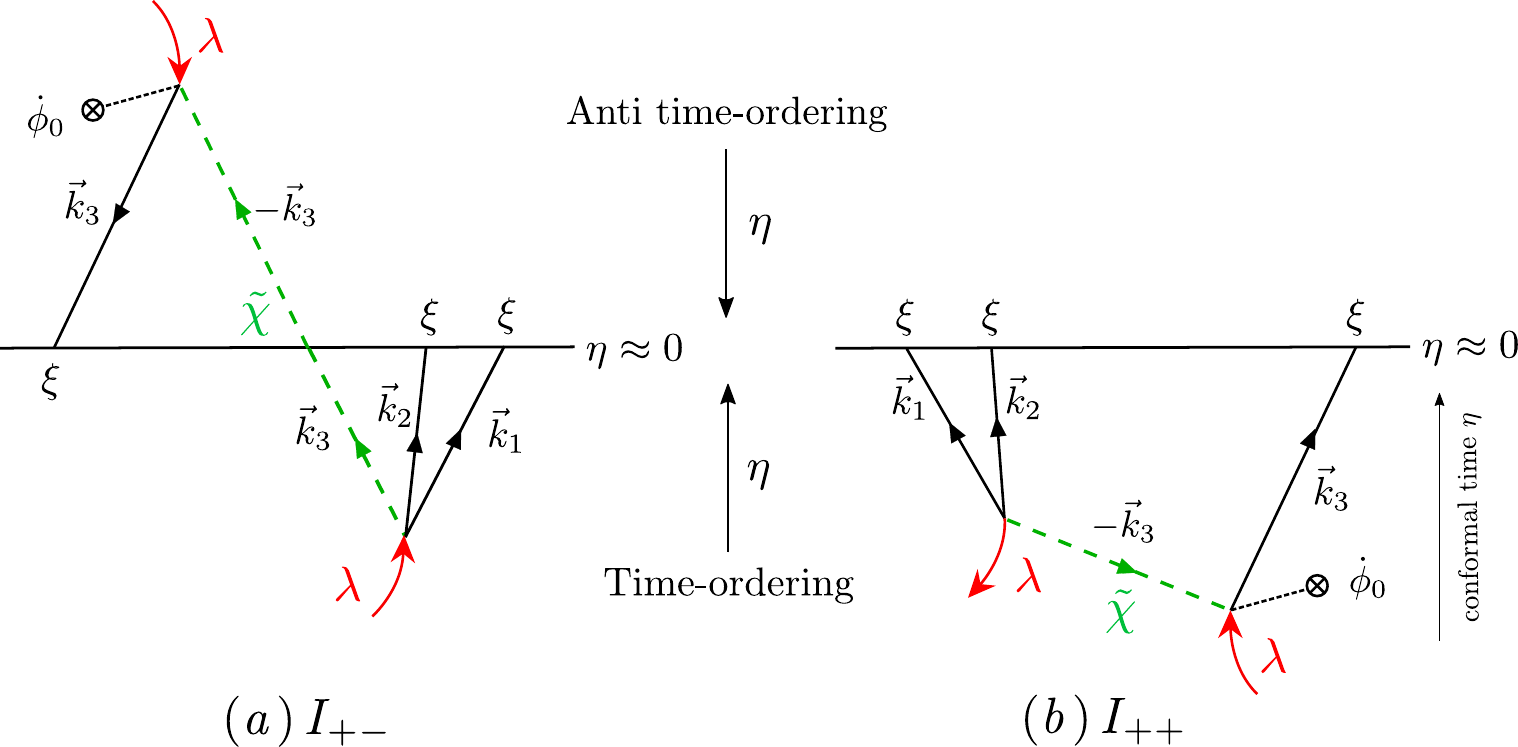}\\			
		\caption{Representation of the dominant tree-level contributions from $I_{+-}$ and $I_{++}$ diagrams to the bispectrum. For the convenience of notation, the time ordered part of the diagram is shown below the $\eta \approx 0$ line with time running upwards, while the anti-time ordered part is shown above with time running downwards. An incoming (outgoing) arrow at a vertex denotes injection (extraction) of energy. This makes it plausible that the $I_{+-}$ contribution is enhanced compared to the $I_{++}$ contribution, since for the former the ``resonant'' particle production through energy injection is efficient at \textit{both} the vertices. Further explanations can be found in the text.}
		\label{in-in_diagram}
\end{figure}

Let us start with the second term with $F_{+-} $ propagator, which is represented diagramatically in \Fig{in-in_diagram} (a). 
There are two integrals, the one with soft inflaton leg $(k_3)$ coming from anti-time ordering, while the other with hard momenta $(k_{1/2})$ coming from time ordering. 
\begin{align}
    I_{+-} &\supset \beta^{*}_{2} \rho_2\, \int \frac{d\eta_1}{\eta_{1}^{4}} \int \frac{d\eta_2}{\eta_{2}^{4}} \left(\frac{\eta_2}{\eta_1}\right)^{-i\lambda} \,   \tilde{d}_{+}(k_1,k_2;\eta_1) d_{-}(k_3;\eta_2)\,   F_{+-}(k_3; \eta_1,\eta_2) = \beta^{*}_{2}\, \rho_2\, I^{(-)}_{k_3} I^{(+)}_{k_{12}} \;,
\end{align}
where $I^{(-)}_{k_{3}}$ is the $\eta_2 $-integral at the mixing vertex from the anti-time ordered part, while $I^{(+)}_{12}$ is the $\eta_1 $-integral from the time-ordered part with cubic interaction. Using mode functions for the inflaton and complex scalar from \cref{chi_modefunc,infl_modefunc}, we get
\begin{align}
    I^{(-)}_{k_3}=   \int_{-\infty}^{0} \frac{d\eta_2}{\eta_{2}^{4}}\,\left(   \frac{(1+ik_3 \eta_{2})e^{-ik_3 \eta_{2}}}{2k_{3}^{3}}\right) (-\eta_{2})^{-i\lambda} 
	  \times \underset{\bar{f}_{k_{3}}(\eta_{2})}{ \underbrace{ \left(N^{*}_{f} \,(-\eta_{2})^{3/2} H^{(1)}_{i\mu}(-k_{3}\eta_{2})\right)}} \;.
\end{align}
For fields with mass greater than the Hubble scale, the rate of expansion of space-time is much smaller as compared to their natural frequency, i.e., $\dot{\omega}/\omega^2 \ll1$. Thus, the dynamics of the mode functions can be effectively captured by the adiabatic modes. For the ease of calculation, let us make the substitution $x= -k_3 \eta_{2} $. Then the full mode function can be approximated by the adiabatic mode as,
\begin{align}
     N^{*}_{f}\, x^{1/2} H^{(1)}_{i\mu}(x) \, \underset{\mu \gg H}{\sim} \, \left. \frac{e^{-i \int \omega(\eta) d\eta}}{\sqrt{2 \omega(\eta)}}\right\vert_{x=-k\eta} = 
    \frac{e^{i\int \sqrt{1+\frac{\mu^2}{x^2} } dx}}{\sqrt{2} \left( 1+ \frac{\mu^2}{x^2}\right)^{1/4}} \;.
\end{align}
The integrand then separates into a slowly-varying polynomial function and a highly oscillatory phase $g(x)$,
\begin{align}\label{eq:Ipm_k3int_statphase}
   I^{(-)}_{k_3} \sim \frac{1}{2 \sqrt{2} k^{3/2-i\lambda}_{3}} \int_{0}^{\infty} dx \, \frac{x^{-5/2}(1-ix)}{(x^2+\mu^2)^{1/4}} \, e^{i g(x)} \;,\\ 
   \text{where} \qquad g(x) = -\lambda \log(x) + x + \int_{x_0}^{x} \sqrt{1+\frac{\mu^2}{x^2} } dx \;.
\end{align}
Here, $x_0$ corresponds to the time at the start of inflation. The dominant contribution comes from the stationary points where $g'(x_{*})=0$, i.e., $-\frac{\lambda}{x_{*} }+ 1+\sqrt{1+\frac{\mu^2}{x_{*}^{2}}} =0 $.  
In this case, there is only one stationary point at $x_{*}=(\lambda^2-\mu^2)/(2\lambda)$. Using this, we get
$g''(x_{*})=4\lambda^3/(\lambda^4-\mu^4) $ for $\lambda>\mu$.
and the integration can be evaluated as
\begin{align}\label{Ipm3_stat}
    I^{(-)}_{k_3} &\sim \frac{1}{2 \sqrt{2} k^{3/2-i\lambda}_{3}} \frac{x_{*}^{-5/2}(1-ix_{*})}{(x_{*}^{2}+\mu^2)^{1/4}} e^{ig(x_{*})} \sqrt{\frac{2\pi}{g''(x_{*})}} \nonumber\\
    & \sim \frac{\sqrt{\pi}}{k^{3/2-i\lambda}_{3}} e^{i\delta(\lambda,\mu)} \lambda^{1/2}  (\lambda^2-\mu^2)^{-1} .
\end{align}
Here we have used the fact that $x_*\gg1$, true for most of the parameter space we are interested in.
The phase factor $\delta(\lambda,\mu)=g(x_*)-\pi/2$ is independent of $k_3$. The additional phase coming from the choice of initial point $x_0$ is arbitrary, but it cancels with the phase from $I^{(+)}_{k_{12}}$ integral. 

The role of the chemical potential is now apparent. The existence of stationary phase indicates on-shell production of particles. 
The phase $(-\eta)^{-i\lambda}$ coming from the chemical potential can be thought of as \emph{providing} energy $\sim \lambda$ into the vertex, which is converted to produce inflaton and complex scalar fluctuations at time $x_{*}$.
If $\lambda < \mu$, there is no stationary point and we will get an exponential suppression\footnote{In the absence of a stationary point for $\lambda < \mu$, we can perform a Wick rotation $x \rightarrow i x$, and carry out the integration in \cref{eq:Ipm_k3int_statphase} explicitly to see the $e^{-\pi (\mu-\lambda)}$ factor. For more details, refer to a similar calculation within the $I_{++} $ contributions in \cref{Ipp_stat}, where such suppression is always present.} 
\begin{align}\label{eq:largeM_suppression}
    I_{+-} \propto e^{-\pi (\mu -\lambda)},
\end{align}
which as $\lambda \rightarrow 0$, becomes the usual Boltzmann-like suppression $e^{-\pi \mu}$ for heavy fields.

Doing a similar calculation for $I^{(+)}_{k_{12}}$ yields,
\begin{align}
    I^{(+)}_{k_{12}}  &= \frac{1}{4k_{1}^3k_{2}^3}  \int_{-\infty}^{0} \frac{d\eta_{1}}{\eta_{1}^{4}}\,\left( (1-ik_1 \eta_{1})(1-ik_2 \eta_{1}) e^{ik_{12}\eta_{1}}\right) (-\eta_{1})^{+i\lambda}
	\times  \underset{f_{k_{3}}(\eta_{1})}{ \underbrace{ \left(N_f \,(-\eta_{1})^{3/2} H^{(2)}_{i\mu}(-k_{3}\eta_{1})\right)}} \nonumber\\
	&\xrightarrow{x= -k_{3}\eta_{1} }   \frac{1}{4 \sqrt{2} k_{1}^3k_{2}^3 k_{3}^{-3/2+i\lambda}} \int_{0}^{\infty} dx \frac{x^{-5/2}(1+ipx/2)^2}{(x^2+\mu^2)^{1/4}} \, e^{i h(x)} \;,
\end{align}
where
\begin{align}
    h(x) =  \lambda \log(x) - px - \int_{x_0}^{x} \sqrt{1+\frac{\mu^2}{x^2} } dx \;.
\end{align}
Stationary phase occurs when $\frac{\lambda}{x_{*}}-p-\sqrt{1+\frac{\mu^2}{x_{*}^{2}}} =0 $. Then in the large $p$ limit, for $\lambda>\mu$, $x_{*} \approx \frac{\lambda-\mu}{p}$, and $h''(x_{*}) \approx -\frac{p^2}{\lambda-\mu}$.  This gives, 
\begin{align}
    I^{(+)}_{k_{12}} \sim \frac{1}{4\sqrt{2} k_{1}^3k_{2}^3 k_{3}^{-3/2+i\lambda}} \, \frac{x_{*}^{-5/2}(1+ipx_{*}/2)^2}{(x_{*}^{2}+\mu^2)^{1/4}} \,  e^{i h(x_{*})} \, \sqrt{\frac{2\pi}{h''(x_{*})}} \;.
\end{align}
Since $x_{*}$ depends on $p$, we get a $p$-dependent phase from $e^{ih(x_{*})}$, 
\begin{align}
    h(x_{*}) &= \left[ \, \underline{\lambda \log(x)}-px - \left.\left\lbrace\sqrt{x^2+\mu^2} + \underline{\mu\log(x)}- \mu \log\left( \mu^2+\mu\sqrt{x^2+\mu^2}\right)\right\rbrace\right\vert_{x_0}^{x} \,\right]_{x_{*} \approx\frac{\lambda-\mu}{p}} .
\end{align}
In the above, we have taken a squeezed limit $p\gg \lambda/\mu \sim 1$, where
this expression can be simplified to separate out the momentum-dependent phase. The underlined terms in the above expression result in a non-analytic momentum dependence of the form $p^{i(\mu-\lambda)}$,
\begin{align}
    e^{ih(x_{*})} &\underset{p\gg 1}{\rightarrow} e^{i\delta'(\lambda,\mu)} p^{i(\mu-\lambda)},
\end{align}
where $\delta'(\lambda,\mu)$ is a $p$-independent phase. The integral can be written as
\begin{align}\label{Ipm12_stat}
    I^{(+)}_{k_{12}} \sim  \frac{-\sqrt{\pi}}{16k_{1}^3k_{2}^3 k_{3}^{-3/2+i\lambda}}\, e^{i\delta'(\lambda,\mu)}\, \mu^{-1/2}\, p^{3/2+ i(\mu-\lambda)} \;.
\end{align}
Putting together \cref{Ipm3_stat,Ipm12_stat}, we get
\begin{align}\label{eq:Ipm_stationary_approx}
    I_{+-} \supset \beta^{*}_2 \rho_{2} I^{(+)}_{k_3} I^{(-)}_{k_{12}} \,\sim \, \frac{\pi \beta^{*}_2 \rho_{2}}{16  k_{1}^3 k_{2}^3}\, \mu^{-1/2} \lambda^{1/2} (\lambda^2-\mu^2)^{-1} \,p^{3/2+i(\mu-\lambda)} \;,
\end{align}
up to a constant phase.

Going back to the first term in \cref{eq:Ipm_expression} with $G_{+-}$, we observe that the calculation can be repeated with the replacement $\lambda \rightarrow -\lambda$.
However, it can be checked that in this case, the stationary point does not exist for either of the two integrals, and the integrand is highly oscillatory in the entire domain of integration $\eta \in (-\infty,0)$. This gives an exponentially suppressed contribution, and hence can be ignored in our estimate.\footnote{For $\lambda<0$, $G_{+-}$ gives the dominant contribution while $F_{+-}$ contribution is exponentially suppressed.}

From this analysis, it becomes clear that the correct way to think of $I_{+-}$ diagrams is \Fig{in-in_diagram} (a). In the figure, we represent anti-time ordered part above the line denoting the end of inflation ($\eta \approx 0$) with time running downwards. This is simply a convenient notation to differentiate between $I_{+-/-+}$ and $I_{++/--}$ diagrams.
The interaction Hamiltonian evolves the initial Bunch-Davies vacuum into a superposition of excited states $|p_{i}...p_{n_{\xi}};q_{j}...q_{n_{\tchi}}\>$ where $n_{\xi}$ is the number of inflaton excitations while $n_{\tchi}$ is the number of excitations of the heavy field. The correlation functions essentially picks the amplitude of the state $|k_{3};k_{3}\>$ from the `bra' and the amplitude of $|k_{1},k_{2};k_{3}\>$ from the `ket'. 

The direction of the arrow for chemical potential $\lambda$ in \Fig{in-in_diagram} depends on the phase. For the vertices from time ordered part, $(-\eta)^{i \lambda}$ corresponds to `injection' of energy into the vertex, while $(-\eta)^{-i \lambda}$ corresponds to `removal' of energy. This is exactly opposite for the anti-time ordered part. 
This will also help us understand why $I_{++}/I_{--}$ contribution is sub-dominant.

The $I_{++}$ diagram is different than $I_{+-}$ in the sense that the produced heavy particle decays into inflaton at a later time as seen in \Fig{in-in_diagram} (b). This means that while the energy of order $\lambda$ is injected at one vertex to boost particle production, similar amount of energy is removed at the other vertex. If $ \lambda >\mu $, the production of massive particle at $ k_{3} $ vertex is efficient, i.e. stationary point exists for the oscillating integrand. But the decay of the produced particle into inflatons at $k_{12}$ vertex is suppressed as the particle has to decay not only into inflatons but also give away energy of order $ \lambda $ which is greater than its own energy $\sim \mu $. This can be seen by the absence of a stationary point for the integral at this vertex, resulting in the suppression of the form $ e^{-\pi(\lambda-\mu)} $. On the other hand, if $ \mu >\lambda $, the reverse is true, that is, decay at $ k_{12} $ will be efficient, but production at $ k_{3} $ will be suppressed as $e^{-\pi (\mu - \lambda)}$. All in all, we expect $I_{++}$ to at least have the suppression of the form $e^{-\pi|\lambda-\mu|}$.

We postpone the calculation of the sub-dominant $I_{++} $ diagram to \cref{Ipp_stat} (the full analytic treatment can be found in \cref{subsec:Ipp_full}), and instead discuss some of our main results in the following sections. 

\subsection{Constraints}\label{Constraints}
Before looking at the strength of the non-analytic signal, we determine the constraints on various parameters of our model in \cref{Lrotated}. To ensure a controlled derivative expansion in $(\partial \phi )^2/\Lambda^{4}$, we require $\Lambda > \dphiv^{1/2}$.
From the Planck data \cite{Akrami:2018odb}, $H^4/\dphiv^2 = 8.25 \times 10^{-8}$ at $k_{*} = 0.05$ Mpc$^{-1}$ for slow-roll inflation. This gives $\dphiv \approx (60H)^2$ implying the chemical potential $\lambda \lesssim 60H$. Thus $\lambda$ can be much larger than the typical energy of the expanding space-time during inflation $\sim H$ while ensuring EFT control.   

We must also ensure that the standard inflaton dynamics is not affected to first order and we have perturbative control over the interactions. The VEV of the complex scalar in \cref{chi0} modifies the kinetic energy of the inflaton as $ (\partial \phi)^2 \rightarrow \left[ 1 - 2(1-c) |\tchi_{0}|^2 / \Lambda^2\right](\partial \phi)^2$. This gives a constraint on the VEV as
\begin{align}\label{Constraint:kin}
   \frac{|\kappa|^2}{\Lambda^2} \ll 1 \;.
\end{align}

We also require the correction to the power spectrum from the interaction Hamiltonian to be sub-leading. 
The dominant corrections come from the symmetry-breaking term $\mathcal{H}_{\text{int}} \supset \alpha \tchi e^{-i \phi/\Lambda} \sim - \alpha \kappa^{\dagger} (\xi/\Lambda)^2 -i \alpha e^{-i \phi_0/\Lambda} \delta\tchi ( \xi/\Lambda) + \text{c.c.}$. These two contributions are comparable, and can be estimated to be $\Delta P_{\mathcal{R}}/P_{\mathcal{R}} \sim \alpha |\kappa|/\Lambda^2$, which should be $\ll 1$. For $\lambda > \meff$, $\kappa \sim \alpha/\lambda^2$. Then it is then easy to see than this constraint is stronger than \cref{Constraint:kin}, and the overall constraint is simply $\alpha / \Lambda \ll \lambda$. In order to have only a percent level correction to the power spectrum \cite{Akrami:2018odb}, we take a more conservative value for the coefficient of the symmetry-breaking term, 
\begin{align}\label{eq:alpha_constraint}
    \frac{\alpha}{\Lambda}= 0.1 \,  \lambda
\end{align}
We will use this value of $\alpha$ while evaluating the strength of NG in the next section.


\subsection{Central results}\label{sec:results}
With the technical calculation out of the way,
we are now in a position to explore the strength of the non-gaussian signal in different parameter regimes. In this subsection, we present the central results of this paper.

Let us estimate the size and the shape of the bispectrum in the squeezed limit using \cref{eq:Fsqueezed_def},
\begin{align}
    &F_{\text{squeezed}} \approx \frac{5}{12} \dphiv \frac{\< \xi_{\vec{k}_1}\xi_{\vec{k}_2}\xi_{\vec{k}_3} \>^\prime}{\< \xi_{\vec{k}_1} \xi_{-\vec{k}_1}\>^\prime\< \xi_{\vec{k}_3} \xi_{-\vec{k}_3}\>^\prime},
\end{align}
where we take 
\begin{align}
    \< \xi_{\vec{k}_2}\xi_{\vec{k}_3}\xi_{\vec{k}_1} \>^\prime \simeq I_{+-} + I_{-+},
\end{align}
since the dominant contribution comes from $I_{+-}$ and $I_{-+}$, as seen previously in \cref{Ipm_stat}. From \cref{eq:Ipm_stationary_approx} and \cref{PowerSpectrum}, we get \footnote{We will drop some $p-$independent phases, since here we will be interested in the magnitude of NG, $f_{\text{oscil}}$ defined below.}
\begin{align}
    F_{\text{squeezed}} \approx  \frac{5}{12} \dphiv  (4 k_{1}^{3}k_{3}^{3})  \frac{ 2\pi \beta^{*}_2 \rho_{2}}{16  k_{1}^3 k_{2}^3}\, \mu^{-1/2} \lambda^{1/2} (\lambda^2-\mu^2)^{-1} \,p^{3/2+i(\mu-\lambda)} +\text{c.c.}.
\end{align}
Here, we have included the symmetry factor of 2 from $k_{12}$ vertex.
Inserting values for $\beta_2, \, \rho_2$ from \cref{Hmix_coeff,Hint3_coeff}, we get the parametric form of NG in squeezed limit $k_{1}\approx k_2 \gg k_3$,
\begin{align}
     F_{\text{squeezed}} \simeq \frac{5\pi}{12\sqrt{2}} \frac{\dphiv}{\Lambda} \,\left(\frac{\alpha}{\Lambda}\right)^{2}\, 
     \left(\frac{\lambda}{\mu}\right)^{1/2}\, (\lambda^2 -\mu^2)^{-1}\, \left(\frac{p}{2}\right)^{-3/2+i(\mu-\lambda)}+\text{c.c.}.
\end{align}
We saturate the constraint on $\alpha$ given in \cref{eq:alpha_constraint} and use $\lambda \equiv \dphiv/\Lambda$ to get
the approximate form of NG in the squeezed limit as
\begin{align}\label{F_stat}
    F_{\text{squeezed}} \simeq \frac{5\pi}{12\sqrt{2}} \times 10^{-2}\times
    \frac{\lambda^{7/2}}{\mu^{1/2} \, (\lambda^2 -\mu^2)}\, \left(\frac{p}{2}\right)^{-3/2+i(\mu-\lambda)}+\text{c.c.}.
\end{align}
Let us denote the amplitude of the non-analytic signal in the bispectrum by $f_{\text{oscil}}$ such that,
\begin{align}
    F_{\text{squeezed}} \approx |f_{\text{oscil}}(\mu,\lambda)| \left(\, \, \left(\frac{p}{2}\right)^{-3/2+i(\mu-\lambda)} +\text{c.c.} \right),
\end{align}
Then, 
\begin{align}\label{eq:fnl_stat}
    |f_{\text{oscil}}(\mu,\lambda)| = \frac{5\pi}{12\sqrt{2}} \times 10^{-2}   \frac{\lambda^{7/2}}{\mu^{1/2} \, (\lambda^2 -\mu^2)}.
\end{align}
From here onward, we will use $|f_{\text{oscil}}|$ to characterize the size of NG.
\begin{figure}[t]
	\centering
		\includegraphics[width=0.65\textwidth]{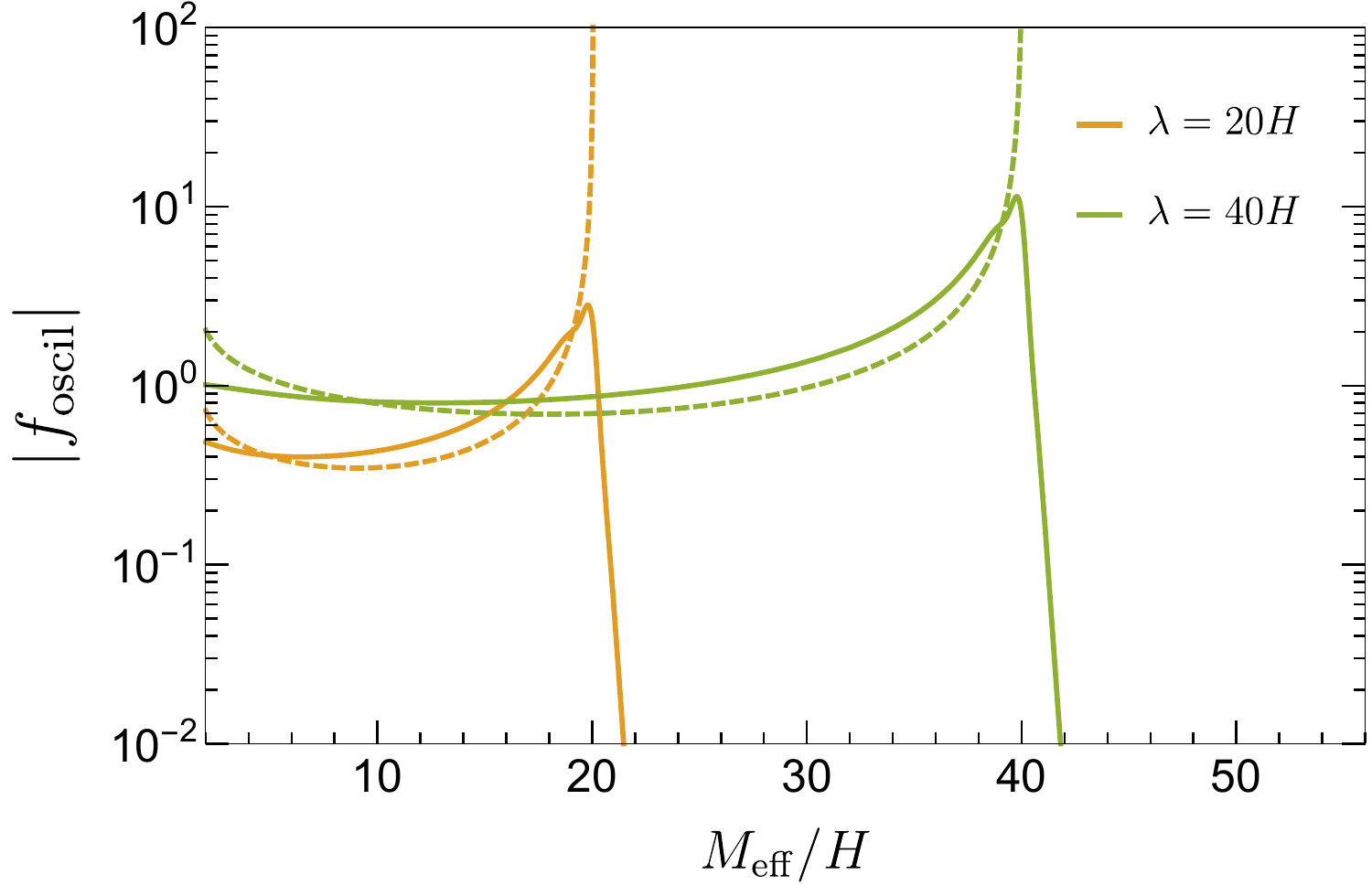}
		\caption{Comparison of $|f_{\text{oscil}}|$ as a function of $\meff$ using stationary phase (dashed) and the full analytic form (solid) for two different values of the chemical potential $\lambda=20H$ (yellow) and $\lambda=40H$ (green), at $c=0.5$ for $\alpha/\Lambda = 0.1 \lambda$. The parametric dependence matches for $\mu < \lambda$ as expected with a difference of an $\mathcal{O}(1)$ factor. This mismatch can be attributed to contributions from other vertices which we did not consider in the stationary phase approximation above just for simplicity. Note the full calculation also smooths out the naive divergence in the stationary phase estimate as $\mu\approx \lambda$ since for such $\mu$ the approximation made in our stationary phase estimate breaks down.} 
		\label{stationary_phase}
\end{figure}

In \Fig{stationary_phase}, we compare above expression from the approximate calculation with the exact calculation from \cref{bispectrum_full} for a generic value of $c=0.5$ and $\lambda = 20,\, 40$.
We see that in the region where the approximations are valid, i.e., $\lambda > \mu$, the parametric form of the non-analytic contribution from the stationary phase analysis is in good agreement with the full analytic form. The small discrepancy can be traced to the contributions from other vertices that we did not take into account, especially the mixing vertex with coefficient $\beta_1$.

\begin{figure}[t]
     
        \centering 
        \includegraphics[width=0.75\textwidth]{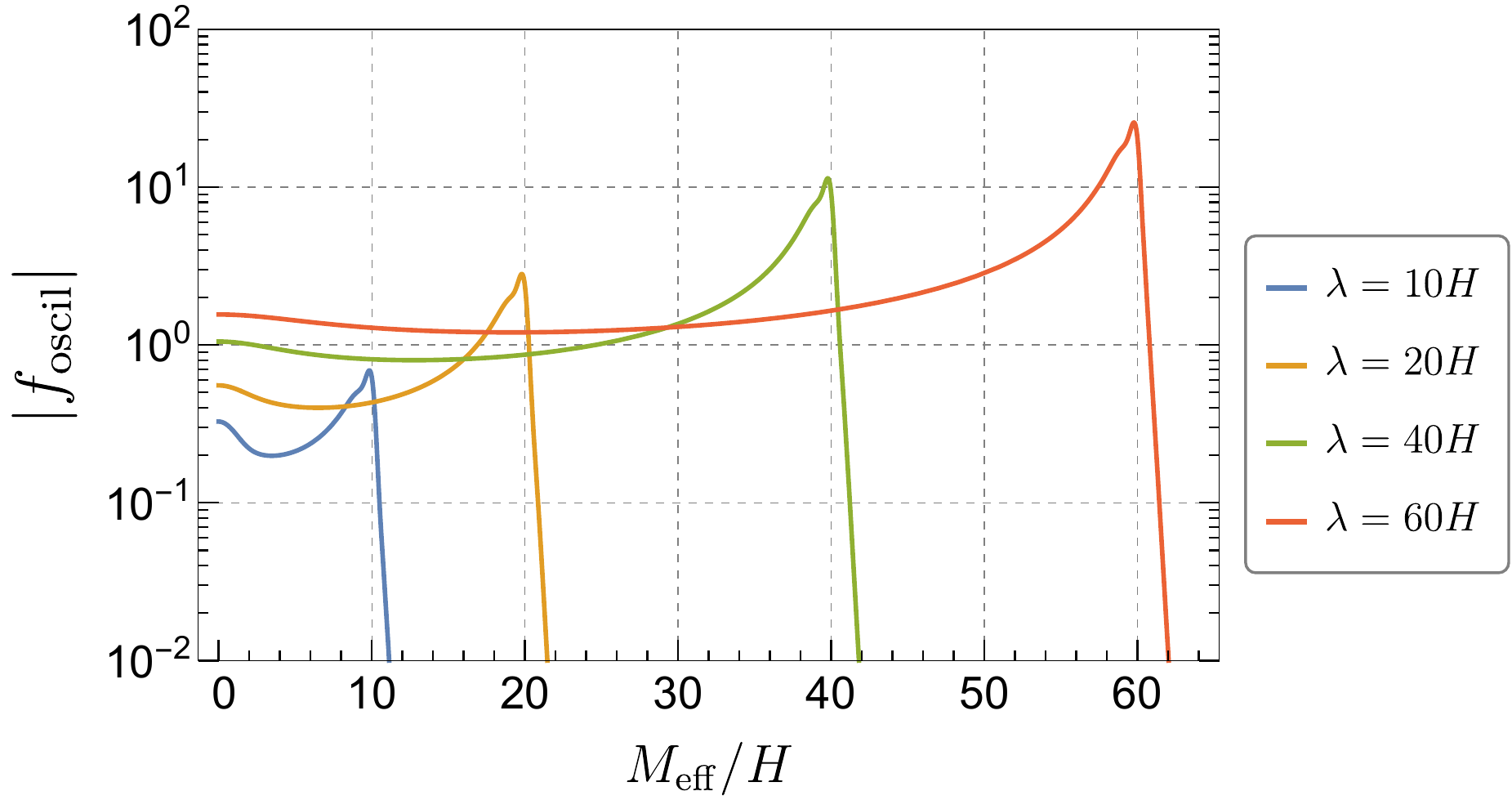}
    \caption{Amplitude of the non-analytic contribution |$f_{\text{oscil}} $| from the full calculation in \cref{bispectrum_full}, plotted against the physical mass $\meff$ of the complex scalar in Hubble units for $c=0.5$ and various values of the chemical potential $\lambda=10H$ (blue), $20H$ (yellow), $40H$ (green), and $60H$ (red). We take $\alpha/\Lambda=0.1\lambda$. The expression for NG has a subdominant dependence on the parameter $c$, and hence we only show a representative plot for a generic value $c=0.5$. Note that the range $0<\meff \lesssim \lambda$ corresponds to $-\lambda^2/2 < M^2 \lesssim \lambda^2/2$ for the ``bare'' mass. The tachyonic regime for $M$ (as opposed to $\meff$), $M^2<0$, corresponds to a post-inflationary instability to generating a significant $\chi$ VEV. During inflation, this instability is absent in the regime of non-tachyonic de Sitter mass $\meff$ plotted. No fine-tuning of parameters is needed to obtain the large signals shown.}
    \label{fnl_lambda_c}
\end{figure}

Having confirmed the validity of our approach, we now study the size of NG for various values of $\lambda$ as shown in \Fig{fnl_lambda_c}. The NG is plotted for a generic value of $c=0.5$, as the plots for other values look similar due to the subdominant dependence of NG on parameter $c$. Some notable features in this plot are:
\begin{itemize}
    \item The most important result is that we obtain an observable NG for a large range of physical masses with a high reach, $H< \meff \lesssim \lambda \,\,(\lesssim 60 H)$.\footnote{By observable, we mean $f_{\text{NL}} \gtrsim O(0.01)$, as explained in \cref{Prelim}.} Note that this corresponds to an equally large range in ``bare'' mass parameter $M^2$ (which dominates after inflation), where for any value of $c$, an usuppressed signal can be obtained for $ -(1-c) \lambda^2<M^2 <c \lambda^2$.
    
    
    \item The non-analytic component of NG has a weak dependence on $\meff$, which gives a plateau-like behavior in $|f_{\text{oscil}}|$ for $\meff < \lambda$. 
    The signal peaks at $\meff \simeq \lambda$ due to the resonance behaviour.
    After that, the exponential suppression kicks in, \cref{eq:largeM_suppression}, and the signal drops as $|f_{\text{oscil}}| \sim e^{-\pi (\mu-\lambda)}$.
    
\end{itemize}
\begin{figure}[t]
	\centering
		\includegraphics[width=0.68\textwidth]{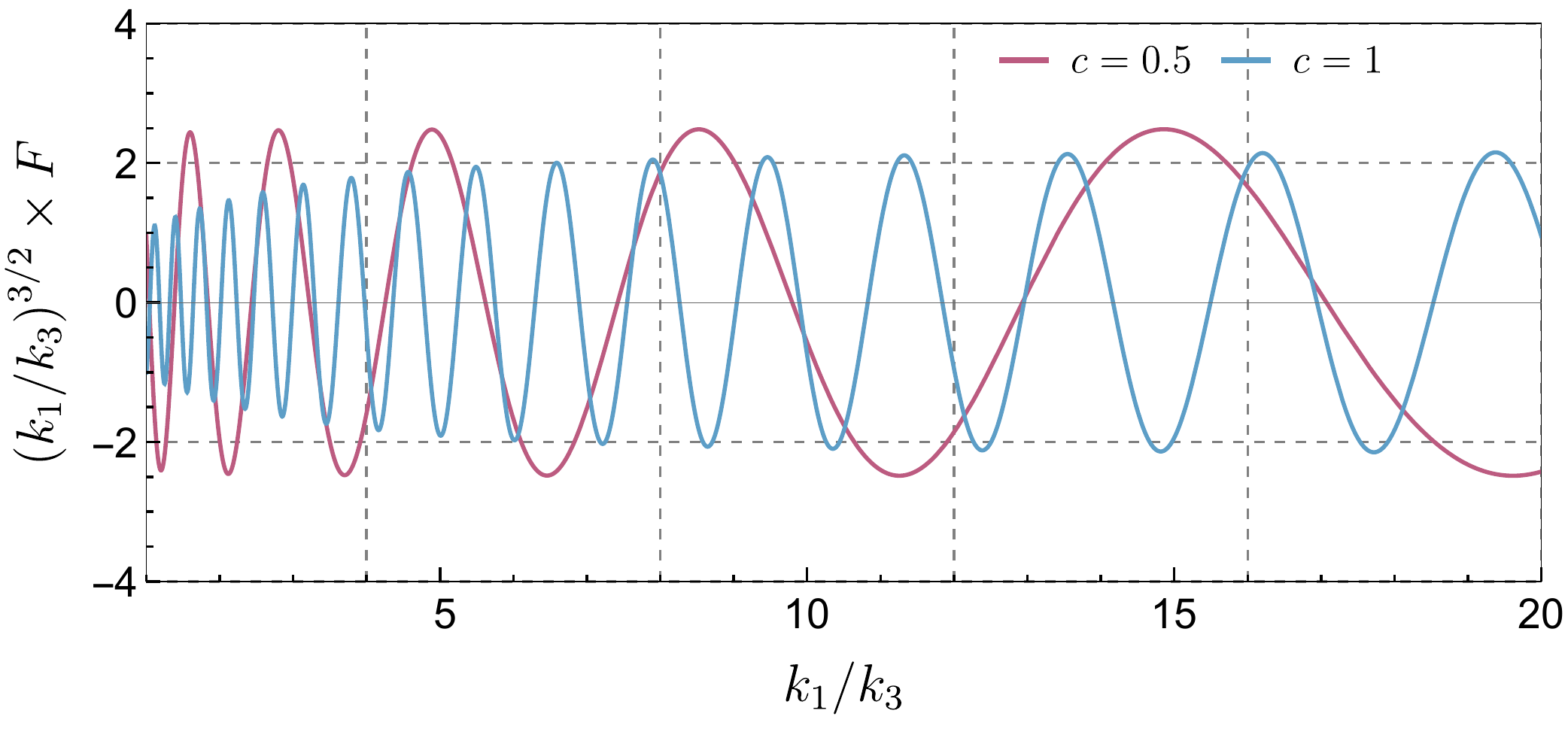}\\	
		\caption{The oscillatory signal as a function of the momentum ratio $k_1/k_3$ for $M=5H$ and $\lambda=40H$.
		The two different curves correspond to  $c=0.5$ or $\meff \simeq 28.7 H$ (purple) and  $c=1$ or $\meff = 5 H$ (blue).} 
		\label{fig:p-dep}
\end{figure}

It is useful to understand how our results for the scalar chemical potential compare to a chemical potential for a heavy fermionic field (see e.g. \cite{Chen:2018xck}). The analogy is closest when we restrict to $c=0$ and $M^2 > 0$, in which case, $\meff = \sqrt{M^2 +\lambda^2} >\lambda$ so that the signal is always exponentially suppressed. We can approximate the suppression factor of \cref{eq:largeM_suppression} for small $M$ as $ e^{-\pi (\mu-\lambda)} \approx e^{-\pi M^2/2\lambda}$, as appears in the case of fermionic chemical potential. We then see that in this regime, the mass range is limited to $0<M \simeq \sqrt{\lambda}$.


We also demonstrate the oscillatory behavior of the NG signal
for benchmark points: $\lambda=40H$ and $\meff = 5H, \,28.7H$ in \Fig{fig:p-dep}. The parameter $c$ affects the frequency of oscillation by modifying $\meff$ as expected.
Also, notice that the amplitude of the oscillations has a small dependence on the ratio $k_1/k_3$, which can be used to infer $\meff$, as we will see in \cref{Sec:Meff_extract}.

In conclusion, we have demonstrated an enhanced NG signal for heavy complex scalar fields with chemical potential for a wide range of parameter space as seen in \Fig{fnl_lambda_c}. The NG is at tree level and hence, the calculation is robust and much simpler than the loop-level processes previously considered in the context of chemical potentials. In \Fig{fig:p-dep}, we see that the signal has a distinct oscillatory signature with many observable oscillations before the $p^{-3/2}$ dilution takes over (this dilution is however factored out on the vertical axis in \Fig{fig:p-dep}).
\subsection{Sub-dominant contribution: $I_{++}$ diagrams}\label{Ipp_stat}
In this section, we go through the calculation of $I_{++}$ diagram to convince the reader that it is indeed exponentially suppressed. Taking the same vertices as for the $I_{+-}$ diagram in \cref{Ipm_stat}, we write the $I_{++}$ contribution
\begin{align}
    I_{++} &\supset (-i) (-i)\int \frac{d\eta_1}{\eta_{1}^{4}} \int \frac{d\eta_2}{\eta_{2}^{4}} \,   \tilde{d}_{+}(k_1,k_2;\eta_1) d_{+}(k_3;\eta_2)\times\, \nonumber\\[0.3em]
    &\hspace{6em} 
    \left\lbrace \beta_2 \rho^{*}_{2}\, \left(\frac{\eta_2}{\eta_1}\right)^{i\lambda}F_{++}(k_3; \eta_1,\eta_2) + \beta^{*}_{2} \rho_2\, \left(\frac{\eta_2}{\eta_1}\right)^{-i\lambda}G_{++}(k_3; \eta_1,\eta_2) \right\rbrace\;.
\end{align}
There is a time-ordering issue in $I_{++}/I_{--}$ diagrams. But the time scales of the dominant contribution are well separated and thus the integrals can be extended to the entire domain of integration. From the previous calculation of $I_{+-}$ diagrams (we will check for consistency later in the calculation), the integral at $k_3$ is dominated by earlier time as compared to the integral at $k_{12}$. This simplifies the time ordering. We focus on the part with $F_{++}$ propagator which corresponds to \Fig{in-in_diagram} (b). 
\begin{align}
    I_{++} &\supset -\beta_2\, \rho_2^{*}\, \< \xi_{0}^{3}\cdot  \left\lbrace \int_{\eta_{1}=0}^{\infty}  \xi^2\,\deltchi^{\dagger}(-\eta_1)^{-i\lambda} \int_{\eta_{2}=0}^{\infty}  \xi \deltchi (-\eta_2)^{i\lambda}) \right\rbrace\>\nonumber\\
    &\sim -\beta_2\, \rho_2^{*}\, \tilde{I}^{(+)}_{k_{12}} \tilde{I}^{(+)}_{k_3},
\end{align}
where tilde denotes integrals in $I_{++}/I_{--}$ diagrams. $\tilde{I}^{(+)}_{k_3}$ is dominated by the contribution from stationary point at $
-k_3\eta_2\equiv x_{*}\sim (\lambda^2-\mu^2)/(2\lambda)$, and it can be evaluated using similar procedure used to get \cref{Ipm3_stat},
\begin{align}\label{Ipp3_stat}
   \tilde{I}^{(+)}_{k_3} \sim \frac{\sqrt{\pi}}{ k^{3/2+i\lambda}_{3}} e^{-i\delta(\lambda,\mu)} \lambda^{1/2}  (\lambda^2-\mu^2)^{-1}.
\end{align}
The integral at $k_{12}$ vertex on the other hand does not have a stationary point in the domain of integration for $\lambda>\mu$.
\begin{align}
     \tilde{I}^{(+)}_{k_{12}} &\sim \frac{1}{4\sqrt{2}k_{1}^3k_{2}^3 k_{3}^{-3/2-i\lambda}} \int_{0}^{\infty} dx \frac{x^{-5/2}(1+ipx/2)^2}{(x^2+\mu^2)^{1/4}} \, e^{i \tilde{h}(x)}\\
    \text{where}\quad \tilde{h}(x) &=  -\lambda \log(x) - px + \int^{x} \sqrt{1+\frac{\mu^2}{x^2} } dx .
\end{align}
Then $\tilde{h}'(x) = -\frac{\lambda}{x}-p+\sqrt{1+\frac{\mu^2}{x^2}}$ is never zero for $x>0$ and $\lambda>\mu$. To evaluate this integral, we instead perform a Wick rotation $x\rightarrow -ix$. Then the phase factor $e^{-ipx}$ becomes an exponential $e^{-px}$ which gives contribution in the region $x<1/p$. This justifies the time-ordering. Since $p>1$, we can  take the late time limit of the adiabatic mode function which goes as $ x^{i\mu}$,
\begin{align}
    \tilde{I}^{(+)}_{k_{12}} &\sim \frac{(-i)^{-3/2}}{4\sqrt{2}k_{1}^3k_{2}^3 k_{3}^{-3/2+i\lambda}}\mu^{-1/2} (-i)^{-i(\lambda-\mu)} \int_{0}^{\infty} dx x^{-5/2}\left(1-\frac{px}{2}\right)^2 x^{-i(\lambda-\mu)} e^{-px} \nonumber\\
    &\sim \frac{(-i)^{-3/2}}{4\sqrt{2}k_{1}^3k_{2}^3 k_{3}^{-3/2+i\lambda}}\left(\frac{p}{2}\right)^2\mu^{-1/2} e^{-\pi(\lambda-\mu)/2}\, \Gamma\left(\frac{1}{2}-i(\lambda-\mu)\right) p^{-1/2+i(\lambda-\mu)}.
\end{align}
Here we have only kept the largest contribution. Taking asymptotic expression for gamma function, $\Gamma(x+iy) \xrightarrow{y\rightarrow \infty} \sqrt{2\pi} |y|^{x-1/2} e^{-\pi |y|/2}$, we simplify above expression 
\begin{align}\label{Ipp12_stat}
    \tilde{I}^{(+)}_{k_{12}} &\sim \frac{\sqrt{\pi}}{16 k_{1}^3k_{2}^3 k_{3}^{-3/2+i\lambda}} \mu^{-1/2} e^{-\pi(\lambda-\mu)}  p^{3/2+i(\lambda-\mu)}.
\end{align}
Combining \cref{Ipp3_stat,Ipp12_stat},
\begin{align}
    I_{++} \supset-\beta_2\, \rho_2^*\, \tilde{I}^{(+)}_{k_{12}} \tilde{I}^{(+)}_{k_3} \sim -\frac{\pi \beta_2 \rho_2^*}{16 k_{1}^3 k_{2}^3}\, \mu^{-1/2} \lambda^{1/2} (\lambda^2-\mu^2)^{-1} \,e^{-\pi(\lambda-\mu)} \,p^{3/2+i(\mu-\lambda)}.
\end{align}
In the regime where $\lambda<\mu$, we still get the suppression $e^{-\pi|\mu-\lambda|}$ but from the $k_{3}$ vertex instead. We see that $I_{++}$ is generally sub-dominant to $I_{+-}$ by a factor of $e^{-\pi|\mu-\lambda|}$, and only contributes significantly in a small region when $\meff$ is close to $\lambda$. 


\subsection{Symmetry breaking mass term}\label{subsec:MassSymmBreak}
Till now, we have considered a linear symmetry-breaking term for the complex scalar. 
However, we can consider a more general derivative coupling of the inflaton to the scalar fields as discussed in \cref{eq:twoScalar_chemPot}, which can be mapped to a quadratic $\chi\chi$-type mass correction. Starting with the  form in \cref{eq:scalar_mix_chemPot},
\begin{align}
    \mathcal{L}_{\rm int} \supset \, -(\lambda_1 \dot{\sigma}_1  \sigma_2 - \lambda_2 \dot{\sigma}_2  \sigma_1 ),
\end{align}
let us write this in terms of a complex scalar $\chi = \sigma_1 + i \sigma_2$,
\begin{align} 
    {\cal L}_{\rm int} \supset  \, \left\lbrace  \frac{i(\lambda_2 +\lambda_1)}{4}\, \left(  \dot{\chi}^{\dagger} \chi - \dot{\chi} \chi^{\dagger} \right) +\frac{i(\lambda_2-\lambda_1)}{4}\, \left(  \dot{\chi}^{\dagger} \chi^{\dagger} - \dot{\chi} \chi \right) \right\rbrace .
\end{align}
Notice that the first term is a chemical potential $\sim \lambda J^{0} $. The second term represents the extent of symmetry-breaking and is $\propto (\lambda_2 -\lambda_1)$ as expected. It can be simplified to a symmetry-breaking mass correction using integration by parts, and then rotating the complex field $\chi \rightarrow e^{-i \pi /4} \chi$ to get rid of the extra phase factors,
\begin{align}
   {\cal L}_{\rm int} \supset   \, \left\lbrace  \frac{i(\lambda_2+\lambda_1)}{4 }\,\left(  \dot{\chi}^{\dagger} \chi - \dot{\chi} \chi^{\dagger} \right) +\frac{3 (\lambda_2-\lambda_1)}{8}\, \left(  \chi^{\dagger} \chi^{\dagger} + \chi \chi \right) \right\rbrace. 
\end{align}
In the presence of a $Z_2$-symmetry $\chi\rightarrow-\chi$, this constitutes the leading $U(1)$-breaking.
It is also roughly analogous to the studies of chemical potential in fermions, which is another motivation to study its effects in greater detail.

The above considerations give rise to a Lagrangian similar to \cref{L} but with a quadratic symmetry-breaking $m^2 \chi^2$ instead,
\begin{align}
    \mathcal{L}_{\chi} = \sqrt{-g} \left\lbrace \right.
    & \left. -|\partial \chi|^2 
		-M^2 |\chi|^2  \right. \nonumber\\[0.3em]
	 & \left. 
	 -\frac{i \partial_{\mu} \phi}{\Lambda} \left(\chi \partial^{\mu} \chi^{\dagger}\, -  \chi^{\dagger}\partial^{\mu} \chi\, \right)
	-  \frac{c\,(\partial \phi)^2 }{\Lambda^{2}} |\chi|^2
	-\frac{m^2}{2}\left(\chi\chi+\chi^\dagger \chi^\dagger\right) \, \right\rbrace \;. 
\end{align}
Then after the usual field redefinition $\chi = e^{-i\phi/\Lambda} \tchi$, we get
\begin{align}
    \mathcal{L}_{\chi} = \sqrt{-g} \left\lbrace   
    \vphantom{\frac{1}{2}} \right.& \left. 
    -|\partial \tchi|^2 -M^2 |\tchi|^2
     \right. \nonumber\\[0.3em]
	 & \left. 
	+ (1-c)\, \frac{(\partial\phi)^2 }{\Lambda^{2}} |\tchi|^2
	-\frac{m^2}{2} \left(\tchi \tchi \, e^{-2i\phi/\Lambda} + \tchi^{\dagger} \tchi^{\dagger} \, e^{2i\phi/\Lambda} \right) \, \right\rbrace .
\end{align}
We work in the regime $H<m \ll \meff $ ($\meff = \sqrt{M^2+(1-c)\lambda^2}$) such that the symmetry-breaking does not lead to explosive particle production. The EOM can be written as,
\begin{align}\label{eq:EOM_MassSymmBreak}
    \partial_{\eta}^{2} \, \tchi - \frac{2}{\eta} \, \partial_{\eta}\tchi + \left( k^2 + \frac{M^{2}_{\text{eff}}}{\eta^2} \right) \tchi +\frac{m^2}{\eta^2} \tchi^{\dagger} (-\eta)^{-2 i \lambda}\,=\,0.
\end{align}
Notice that the EOM for $\chi$ and $\chi^{\dagger}$ are coupled due to the symmetry-breaking mass correction. The diagonalization of non-derivative terms will involve a time-dependent unitary matrix with factors $\propto (-\eta)^{\pm i \lambda}$ from the symmetry-breaking mass correction. This will induce off-diagonal elements from the derivative terms in EOM. Invariably, factors of $(\eta)^{\pm i \lambda}$ can not be removed, which is the source for particle production.

\Eq{eq:EOM_MassSymmBreak} is similar to narrow parametric resonance with a high frequency mass correction. To get some intuition about the effect of $m$, we can consider a simplified EOM for a real scalar field $\sigma$ motivated by \cref{eq:EOM_MassSymmBreak},
\begin{align}
\partial_{\eta}^{2} \, \sigma - \frac{2}{\eta} \, \partial_{\eta}\sigma + \left( k^2 + \frac{M^{2}_{\text{eff}}}{\eta^2} \right) \sigma +\frac{m^2}{\eta^2} \cos(2\lambda t)\sigma\,=\,0.  
\end{align}
Given that we will focus on a regime $\lambda> \meff \gg m > H$, we will treat the expansion of spacetime adiabatically compared to particle production.
We expect a resonance-like behavior when the frequency of the source $ \sim 2\lambda$ is twice the natural frequency of the complex scalar, i.e., when $\omega(\eta_{*}) \sim \lambda/\eta_{*}$ in conformal time. 
Then the condition for resonance is,
\begin{align}
    \sqrt{\frac{k^2}{a_{\text{res}}^{2}}+\meff^2} = \lambda,
\end{align}
and it lasts for the duration when the frequency is inside a resonance band. We only consider the first resonance band, where the amplification is the largest \cite{Kofman:1997yn}:
\begin{align}
    \lambda \left(1-\frac{m^2}{2\lambda^2}\right) \leq \sqrt{k^2/a^2 + \meff^2} \leq \lambda \left(1+\frac{m^2}{2\lambda^2}\right).
\end{align}
This gives time interval of resonance to be,
\begin{align}
    \delta t \sim \frac{m^2}{(\lambda^2-\meff^2)}.
\end{align}
Then the amplification factor can be evaluated to be
\begin{align}\label{eq:MassSymmBreak_ampl_incr}
    e^{\frac{m^4}{\lambda (\lambda^2-\meff^2)}} \sim 1+ \mathcal{O}(m^4/\lambda^3).
\end{align}
We see that the resonance occurs for a very short duration and hence the amplification is small $\sim \mathcal{O}(m^4/\lambda^3)$. This is expected since the physical momentum dilutes exponentially fast.
Above analysis indicates that if we work in the regime where the amplification is small, i.e., $m^4/ \lambda^3 \ll 1$, the symmetry-breaking term can be considered as a perturbation. We can then treat it as a Feynman vertex in the ``insertion'' approximation, while the propagators of the free fields are unchanged. In the following, we estimate its contribution to the bispectrum.
\begin{figure}[t]
	\centering
		\includegraphics[width=0.65\textwidth]{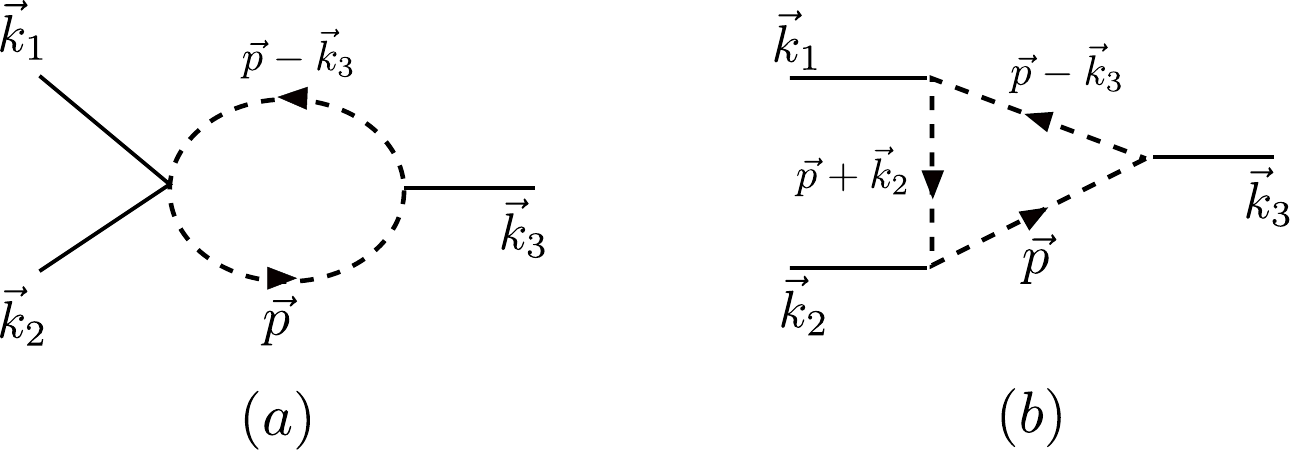}\\	
		\caption{Loop-level contribution to the bispectrum from $m^2 \tchi^2$-type symmetry breaking. The solid lines correspond to inflaton propagators, while dashed lines to complex scalar propagators.} 
		\label{fig:m2_loops}
\end{figure}

The contributions to the bispectrum come from diagrams in \Fig{fig:m2_loops}. The full loop calculation is generally quite challenging, but it can be simplified in certain regimes. In the squeezed limit, the non-analytic signature is prominent when both $\tchi$-modes at the $k_3$ vertex are soft $\sim k_3$. Then the loop integral can be approximated as
\begin{align}
    \int d^3 p \, \approx \, k_3^3.
\end{align}
With this simplification, the contribution to the bispectrum can be estimated using similar methods as in \cref{Ipm_stat} (except now we have two $\tchi$ modes at each vertex and an explicit time-dependence of the form $(-\eta)^{\pm 2 i \lambda}$),
\begin{align}
    F_{\rm squeezed} \sim \frac{1}{ 16\pi^2} \frac{\dphiv}{\Lambda} \left(\frac{m^2}{\Lambda}\right)^2 \lambda\, \mu^{-1} p^{-3+2i(\mu-\lambda)},
\end{align}
where we have included the loop factor $1/16\pi^2$.
For $\Lambda , \mu \sim \mathcal{O}(\lambda)$, this expression can be simplified as
\begin{align}
    F_{\rm squeezed} \sim \frac{1}{16 \pi^2} \left(\frac{m^4}{\lambda^3}\right) \lambda^2 \,p^{-3+2i(\mu-\lambda)}.
\end{align}
Notice that here the non-analytic signal dilutes faster $\sim p^{-3}$ as compared to the case of symmetry-breaking tadpole term, while the non-analytic exponent is twice as large. This is the feature of the contribution being at loop level. The interaction will also modify the power spectrum, which can be expected to be $\sim \mathcal{O}(m^4/\lambda^3)$. Then to ensure only a percent level correction to the power spectrum, we require $(m^4/ \lambda^3) \lesssim 10^{-2}$. Using this constraint, $|f_{\rm oscil}|$ can be estimated as,
\begin{align}
    |f_{\rm oscil}| \sim 10^{-2} \frac{\lambda^2}{16 \pi^2} \xrightarrow{\lambda \sim (10-60) H} \mathcal{O}(10^{-2}- 10^{-1}).
\end{align}
We see that the non-analytic contribution from $m^2\chi^2$-type symmetry breaking term is generically somewhat smaller compared to the case of linear symmetry-breaking model as discussed in \cref{sec:results}. 
While the contribution is small, it could still be within the sensitivity of future experiments. Also, this could be the sole contribution if an additional $Z_2$-symmetry is imposed on the complex scalar field. 
One may also consider the parameter regime where the amplification is large and must be included at leading order. We leave a more detailed study of this model to future work.

\section{Inferring effective mass from the bispectrum}\label{Sec:Meff_extract}
In the cosmological collider physics program, the non-analytic exponent of $p \,(= \, k_{12}/k_3)$ in the squeezed limit allows spectroscopy of heavy fields as seen in \cref{eq:bisprectrum_general}. In our case, the dominant contribution to NG has a non-analytic behavior of the form $\sim p^{-3/2 \pm i (\mu-\lambda)}$ as seen in \cref{F_stat}, which means an independent measurement of $\lambda$ is needed to infer the effective mass of the heavy field. 
In this section, we outline one such procedure which can be used to effectively extract the effective mass and the chemical potential in a  certain parameter regime by observing the change in the non-analytic exponent as a function of the squeezing parameter $p$.

Till now, we have been taking the squeezed limit where $p \gg 1$, but there is another interesting region when $\lambda \gg p, \, \mu$. To get the parametric form of the bispectrum in this regime, we will have to use the following asymptotic form of the hypergeometric function \cite{temme2003large},
\begin{align}
    _{2}F_{1}\left( a+\gamma,b+\gamma,c+\gamma;z\right) \xrightarrow[]{\gamma \rightarrow \infty}\, \frac{\Gamma(c+\gamma)}{\Gamma(b+\gamma)}  \left(\sum_{s=0}^{\infty} f_{s}(z)\, \dfrac{(b)_{s}}{\gamma^{c-b+s}}\right)  (1-z)^{(c-a-b-\gamma)} \label{hyperG_lambda}.
\end{align}
Using this asymptotic form in the analytic expression for the bispectrum (see \cref{bispectrum_full} for details), we can get the non-analytic dependence on $p$ in the large $\lambda$ limit,
\begin{align}
    _{2}F_{1}\left(\underbrace{r\pm i\mu}_{a} \pm i\lambda,\underbrace{r\mp i\mu}_{b}\pm i\lambda,\underbrace{r+1/2}_{c} \pm i\lambda;(1-p)/2\right)
    \overset{\lambda\rightarrow \infty}{\propto}
    \left(\frac{1+p}{2}\right)^{1/2-r\mp i\lambda}\,,
\end{align}
where $r$ is some rational number that depends on the specific interaction. The takeaway is that the asymptotic form in the limit $\lambda \gg p,\, \mu$ is different than the squeezed limit, and the non-analyticity has the form $(1+p)^{\pm i\lambda}$. 

This can also be seen from our approximate stationary phase analysis from earlier. For $\lambda > \mu p$, the production at $k_{12}$ vertex occurs much earlier when the frequency of the complex scalar is dominated by the physical momentum rather than its mass. Hence, the non-analytic part can not capture the information about $\meff$, and hence, it is expected to only depend on $\lambda$.

\begin{figure}[t]
    \centering
    \begin{subfigure}[]{0.52\textwidth}
        \centering
        \includegraphics[width=\textwidth]{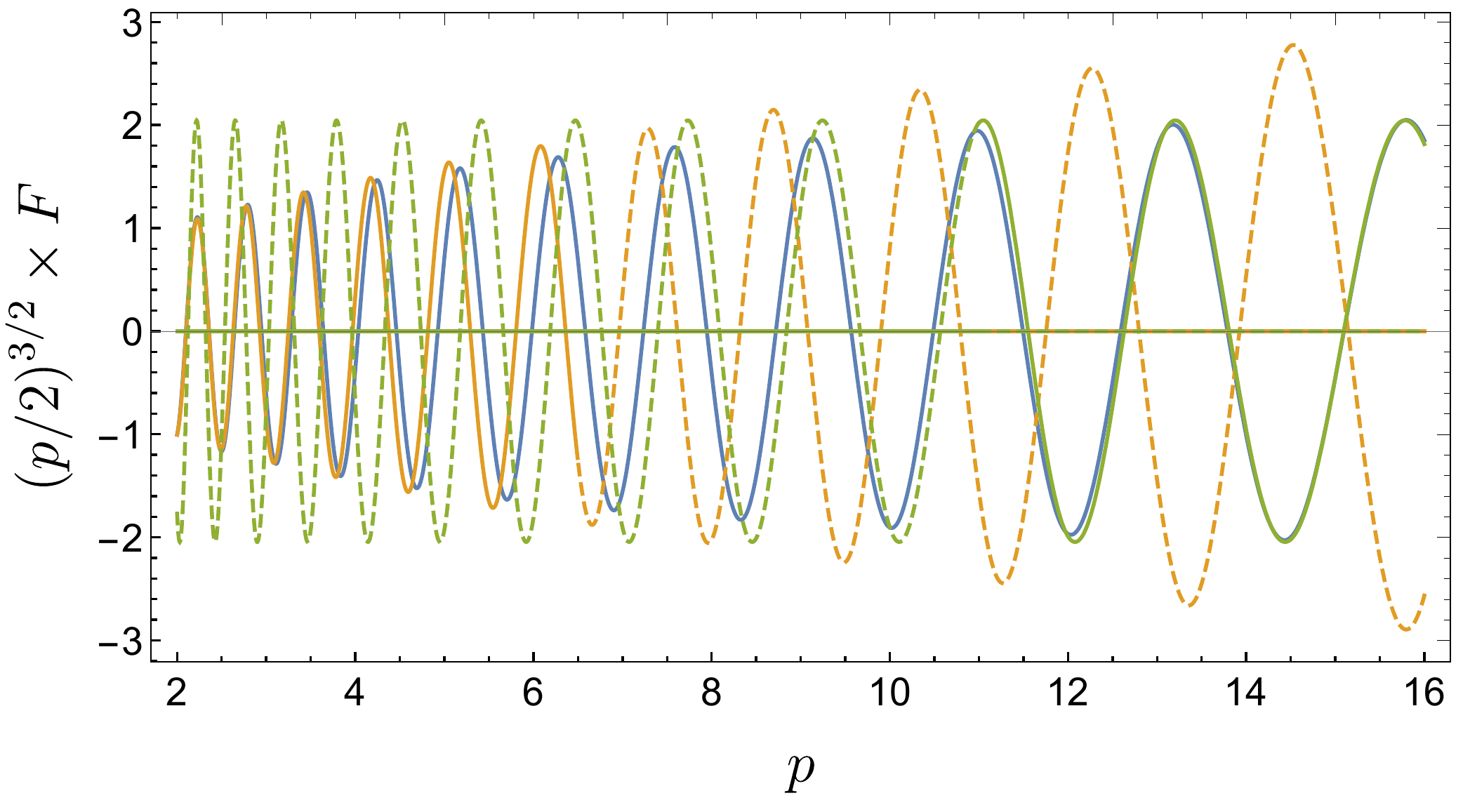}
        \caption{}
    \end{subfigure}
   \quad
    \begin{subfigure}[]{0.43\textwidth}  
        \centering 
        \includegraphics[width=\textwidth]{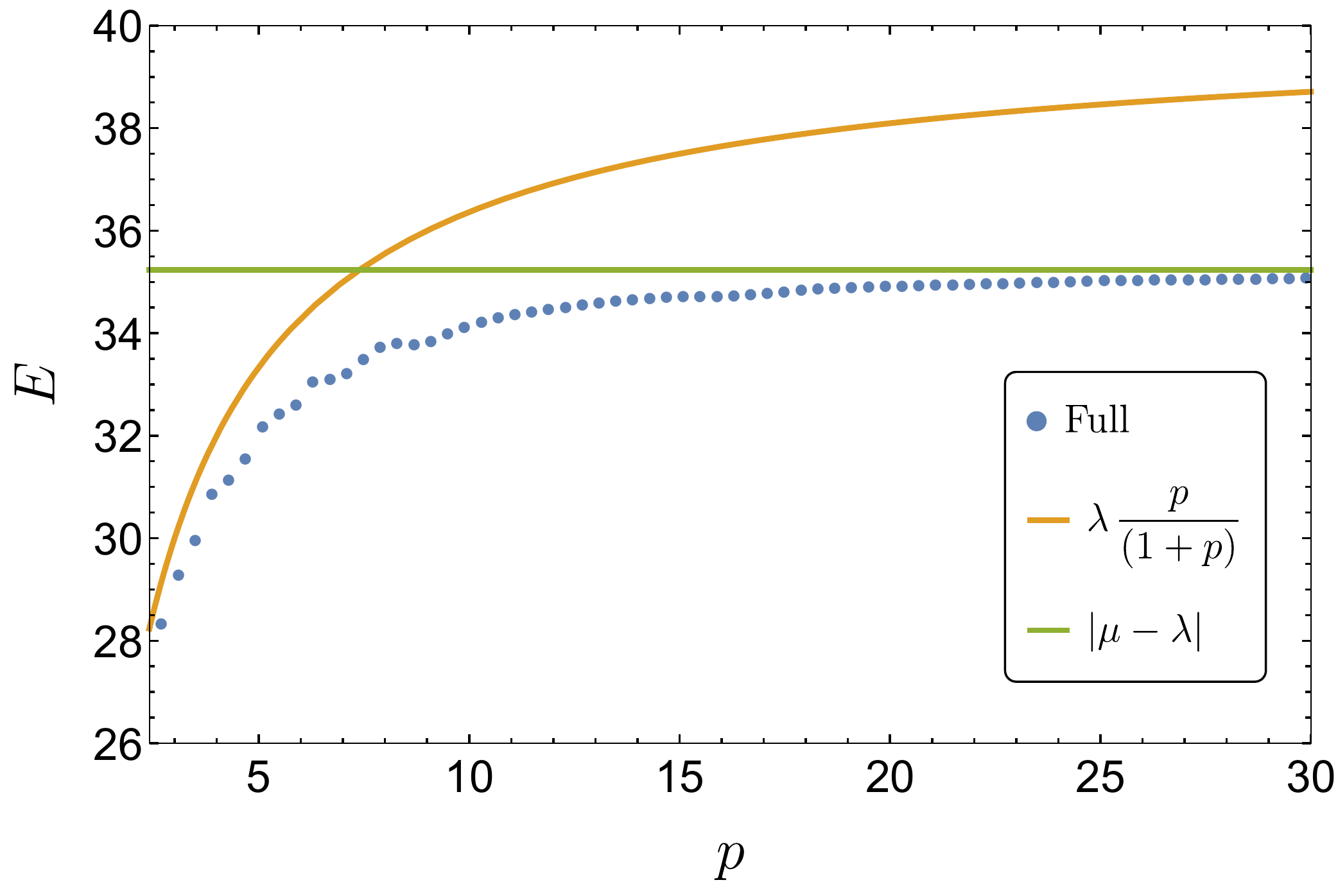}
        \caption{}
    \end{subfigure}
    \caption{Comparison of the full analytic form evaluated in \cref{bispectrum_full} (blue) to the large $\lambda$ approximation (yellow) and large $p$ approximation (green) for $\meff = 5H$ and $\lambda=40H$, using (a) the oscillatory form of the dimensionless NG function $F$ and (b) the non-analytic exponent $E$. We clearly see that as the degree of squeezing is changed the full bispectrum switches between the above two approximations.}
    \label{fig:M_extract}
\end{figure}

In summary, when $\lambda > \mu p$, we expect the non-analytic dependence $\sim (1+p)^{\pm i \lambda}$, and in large $p$ limit, i.e., $\lambda < \mu p$, we expect the non-analyticity of the form $\sim p^{\pm i (\mu-\lambda)}$. In \Fig{fig:M_extract} (a), we see that the full analytic form (blue) agrees with the large $\lambda$ approximation (yellow) for small values of $p$. In the squeezed limit on the other hand, the analytic form matches the large $p$ approximation (green). This corroborates our expectation from the previous discussion.

To extract the effective mass, we can look at the non-analytic exponent as a function of the degree of squeezing. For this purpose, we define a function $E$ such that
\begin{align}
    E &= \text{envelope of}~(p)\partial_p\left(p^{3/2}F(k_1,k_2,k_3)\right) / \text{envelope of}~\left(p^{3/2}F(k_1,k_2,k_3)\right)\\
     &= 
  \begin{cases} 
   \lambda \, \dfrac{p}{(1+p)} & \text{if } \lambda > \mu p \\
   |\mu-\lambda|      & \text{if } \lambda < \mu p.
  \end{cases}
\end{align}
Then $E$ is effectively the non-analytic exponent. From \Fig{fig:M_extract} (b), we see that it also agrees well with our expectation.
Using this feature, we can extract $\mu = \sqrt{\meff^2-9/4}$ (and hence $\meff$) as well as the chemical potential $\lambda$. This procedure is more efficient for $c$ close to 1, i.e, the case of pure chemical potential. It becomes harder to extract the effective mass if $c \ll 1$, when $\mu\sim O(\lambda)$, and we are effectively always in the large $p$ limit.	

\section{Conclusions}\label{sec:conclusions}

In this paper, 
 we have shown that a derivative coupling of the inflaton to heavy scalar fields of ``chemical potential'' type can lead to unsuppressed production of heavy particles and leave observable imprints in primordial non-Gaussianities.
 In particular, the chemical potential can overturn the 
 Boltzmann suppression for masses far above the inflationary Hubble scale $H$, ordinarily present in cosmological particle production. 
 This effect is quite general, requiring only two (or more) real scalar fields where the symmetry rotating one into the other is broken. 
In essence, the unsuppressed particle production above $H$ draws on the large kinetic energy of the inflaton background, $\dot{\phi}_0 \sim (60H)^2$. 
A heavy mass reach up to $\sim 60 H$ can thereby be accomplished while still remaining within a controlled derivative expansion for the inflationary 
 effective field theory.  
 Signatures of such heavy masses could be observed in the bispectrum with strength $\fnl \sim \mathcal{O}(0.01 - 10)$, within the sensitivity of upcoming LSS and future 21-cm experiments.
 
The main results of our analysis are summarized in \Fig{fnl_lambda_c} and \Fig{fig:p-dep}. In our model, the contribution to the bispectrum is at tree-level, which removes the loop-suppression seen in earlier spin-1/2 and spin-1 chemical potential examples.  Importantly, it also makes the calculations analytically tractable and physically transparent. We outlined a procedure to infer the effective mass of the heavy field, which relies on the variation in the non-analytic exponent of the co-moving momentum ratio (degree of ``squeezing'') as in \Fig{fig:M_extract}. 

It is straightforward to generalize our mechanism to more than one heavy complex scalar, each with its own distinct chemical potential $\lambda$, thereby allowing us to explore multiple ``channels''. 
More non-trivially, our mechanism can be extended to massive complex fields with arbitrary non-zero spins. The mechanism can also be extended to the curvaton paradigm \cite{Enqvist:2001zp,Lyth:2001nq,Moroi:2001ct} for inflation, which can exhibit new features. We will take up these generalizations in future work.

\section*{Acknowledgments}
We are grateful to Xingang Chen, Daniel Green, 
Anson Hook, Enrico Pajer 
and Zhong-Zhi Xianyu for helpful discussions. This work is supported by the NSF grant PHY-1914731 and by the Maryland Center for Fundamental Physics (MCFP). SK is also supported in part by the NSF grant PHY-1915314 and the U.S. DOE Contract DE-AC02-05CH11231.
\appendix
 
\section{Important integrals involving Hankel functions}\label{sec:HankelInt}
In this section, we outline the derivation of some of the Hankel integrals that are useful for the calculation of NG. We need to evaluate integrals of the form,
\begin{align}
    \intl{0}{\infty} dx\, x^{n} e^{\mp ip x} \, e^{\mp \pi \mu} H^{(1/2)}_{i\mu} (x) \;.
\end{align}
While the expressions for these Hankel integrals can be found for example in \cite{Kumar:2017ecc}, the derivation is quite instructive (similar treatment can be found in \cite{Arkani-Hamed:2015bza}). We show that it is easily generalised to the imaginary values of $n$, which is relevant for us. We sketch the derivation for integrals with $H^{(2)}$, while a similar procedure can be used to get integrals involving $H^{(1)}$.
\begin{align}
    I_{2}(n,p)= \int_{0}^{\infty} dx \, x^{n} e^{i p x} e^{\pi\mu/2} H^{(2)}_{i\mu}(x)  = \int_{0}^{\infty} dx \, x^{n-3/2} e^{i p x}  f(x) \;,
\end{align}
where $f(x) = e^{\pi \mu/2} \,x^{3/2} H^{(2)}_{i\mu}(x)$ is a solution to the EOM for mode functions as seen in \cref{modeFunc}. Then,
\begin{align}
     & \left[ \partial^{2}_{x} - \frac{2}{x} \partial_{x} + (1+\frac{M^2}{x^2}) \right] f(x) =0 \nonumber\\
    \implies \quad & \int_{0}^{\infty} x^{2+n-3/2} e^{i p x} 
    \left[ \partial^{2}_{x} - \frac{2}{x} \partial_{x} + (1+\frac{M^2}{x^2}) \right] f(x) =0 \;.
\end{align}
From here, we can systematically move the integration past the differential operator using integration by parts to get a differential equation for $I_{2}$ with respect to variable $p$. To clarify the point, consider the last term,
\begin{align}
  \int_{0}^{\infty} x^{2+n-3/2} e^{i p x} \left( 1+ \frac{M^2}{x^2} \right) f(x) 
    = &  \int_{0}^{\infty}  x^{2+n} e^{i p x} e^{\pi\mu/2} H^{(2)}_{i\mu}(x) + M^2 \int_{0}^{\infty} x^{n} e^{i p x} e^{\pi\mu/2} H^{(2)}_{i\mu}(x) \nonumber\\
    = & (-\partial_{p}^{2}+M^2 )I_{2}(n,p) \;.
\end{align}
Similarly for the derivative terms, we integrate by parts. For example, the second term can be written as,
\begin{align}
    & \int_{0}^{\infty} x^{2+n-3/2} e^{i p x} 
   \left(  - \frac{2}{x} \partial_{x} \right) f(x) \nonumber\\
   = & -2 \left\lbrace \int \partial_{x} \left( x^{1+n-3/2}e^{i p x} f(x) \right) - \partial_{x} \left( x^{1+n-3/2} e^{i p x}\right) f(x)\right\rbrace \;.
\end{align}
The first term is the boundary term which can be dropped. Then we are left with,
\begin{align}
     \int_{0}^{\infty} x^{2+n-3/2} e^{i p x} 
   \left(  - \frac{2}{x} \partial_{x} \right) f(x)
   = \left[ 2(n-1/2) + 2ip (-i\partial_{p})\right] I_{2}(n,p) \;.
\end{align}
Following this procedure, we get a differential equation for $I_{2}(n,p)$ as,
\begin{align}
     \left(\, (p^2-1) \partial_{p}^{2}+ 2p\left( n+ \frac{3}{2}\right) \partial_{p} + \left[ M^2 + \left( n- \frac{1}{2}\right) \left( n+ \frac{5}{2}\right) \right] \, \right) \ I_{2}(n,p)=0 \;.
\end{align}
We want the solution to be regular at $p=1$. This implies,
\begin{align}
    I_{2}(n,p) \propto \, _{2}F_{1}\left(n+1+i\mu,n+1-i\mu,n+3/2;\frac{(1-p)}{2}\right)\;.
\end{align}
The normalization is fixed by explicitly carrying out the integration in large $p$ limit using small $x$ expansion of the Hankel function, and then matching it to the large $p$ expansion of hypergeometric function. We also note that this derivation holds even if $n$ has an imaginary part. \\
In summary:
\begin{align}
	&I_{1}(n,p) =e^{-\pi \mu /2} \int_{0}^{\infty}dx \, x^{n} e^{+ipx}H^{(1)}_{i\mu}(x) = \nonumber\\
	&\frac{(i/2)^n}{\sqrt{\pi}\,\Gamma(n+3/2)}\Gamma(n+1+i\mu)\Gamma(n+1-i\mu) _{2}F_{1}(n+1-i\mu,n+1+i\mu,n+3/2,(1-p)/2) \label{H1_int} \;,\\
	& I_{2}(n,p)= e^{\pi \mu /2} \int_{0}^{\infty}dx \, x^{n} e^{-ipx}H^{(2)}_{i\mu}(x)  \nonumber\\
	&= \frac{(-i/2)^n}{\sqrt{\pi}\,\Gamma(n+3/2)}\Gamma(n+1+i\mu)\Gamma(n+1-i\mu) _{2}F_{1}(n+1+i\mu,n+1-i\mu,n+3/2,(1-p)/2) \label{H2_int} \;.
\end{align}

\section{Full calculation of the bispectrum}\label{bispectrum_full}
Our interaction Hamiltonian of the following form,
\begin{align}
    \mathcal{H}_{\text{mix}} \,&=\, \sqrt{-g}\, (-\eta)^{i\lambda}\,\left( \, \beta_{1}  \,\dot{\xi}\,  \tchi \plus \beta_{2} \, \xi\, \tchi \, \right) \plus \text{c.c.} \quad \text{and} \\
    \mathcal{H}_{3} \,&=\,\sqrt{-g}\,(-\eta)^{i\lambda} \, \left( \,  \rho_{1}  \, (\partial \xi)^2\, \tchi \plus \rho_{2} \, \xi\, \xi\, \tchi \, \right) \plus \text{c.c.}.
\end{align}
	\subsection{Full analytic calculation of $ I_{+-/-+} $ diagrams}\label{subsec:Ipm_full}
	    \begin{align}
	    I_{+-} \sim  \< 0|\int \mathcal{H}_{\text{mix}}\,\,\, \xi_0^3 \, \int \mathcal{H}_{3}\, |0 \> .
	    \end{align}
	    There are in all 8 different sub-diagrams contributing to $I_{+-}$: 2 different vertices in both $\mathcal{H}_{\text{mix}}$ and $\mathcal{H}_{\text{int}}$, and two possible contractions of the complex scalar, $\< \tchi \tchi^{\dagger} \>$ and $\< \tchi^{\dagger} \tchi \>$. The contribution from $\< \tchi \tchi^{\dagger} \>$ sub-diagrams are related to those with $\< \tchi^{\dagger} \tchi \>$ by a replacement $\lambda \rightarrow -\lambda$.     
        Thus, it is enough to evaluate terms with contraction $\< \tchi^{\dagger} \tchi \>$ explicitly.
        	\begin{align}
		 I_{+-} \supset  \, \int_{-\infty}^{0}\, \frac{d\eta_{1}}{\eta^{4}_{1}}\, \frac{d\eta_{2}}{\eta^{4}_{2}}\,\langle \, (\beta^*_1 \dot{\xi}+ \beta^*_2 \xi)\, \tchi^{\dagger}\, (-\eta_1)^{-i\lambda}\, \cdot \, \xi_{0} \xi_{0} \xi_{0}\, \cdot \, (\rho_1 (\partial \xi)^2+\rho_2 \xi^2) \, \tchi\, (-\eta_2)^{i\lambda} \, \rangle.
		\end{align}
		Using
		\begin{align}
			\< \dot{\xi}(\eta,\vec{k}) \xi (\eta\rightarrow 0,-\vec{k}) \> = -\frac{\eta^2k^2}{2k^3}\, e^{-ik\eta} \quad \text{and} \quad \< \xi(\eta,\vec{k}) \xi (\eta\rightarrow 0,-\vec{k}) \> = \frac{(1+ik\eta)e^{-ik\eta}}{2k^3},
		\end{align}
		we can write the anti-time ordered contribution at $ k_{3} $ vertex (up to pure phase) as,
		\begin{align}\label{eq:Ipm3_full}
			I_{k_{3}}^{(-)} &= (+i)  \int_{-\infty}^{0} \frac{d\eta}{\eta^{4}}\,\left( -\beta^*_1 \frac{\eta^2k_{3}^2\,e^{-ik_{3}\eta}}{2k_{3}^3} + \beta^*_2 \frac{(1+ik_3 \eta)e^{-ik_3 \eta}}{2k_{3}^{3}}\right) (-\eta)^{-i\lambda} \nonumber \\[0.4em]
			& \qquad \qquad \qquad \qquad \times \underset{\bar{f}_{k_{3}}(\eta)}{ \underbrace{ \left(N_{f}^{*}(-\eta)^{3/2} H^{(1)}_{i\mu}(-k_{3}\eta)\right)}} \nonumber\\
			&= \frac{i\sqrt{\pi}}{4} \frac{1}{k_{3}^{3/2-i\lambda}}\left[ -\beta^*_1 I_{1}\left(-\frac{1}{2}-i\lambda,1\right) +\beta^*_2 \left\lbrace I_{1}\left(-\frac{5}{2}-i\lambda,1\right)-i I_{1}\left(-\frac{3}{2}-i\lambda,1\right) \right\rbrace \right].
		\end{align}
		Similarly, we calculate time ordered contribution from $ k_{12} $ vertex as, 
		\begin{align}\label{eq:Ipm12_full}
			I_{k_{12}}^{(+)} &= \frac{-i}{4k_{1}^3k_{2}^3}  \int_{-\infty}^{0} \frac{d\eta}{\eta^{4}}\,\left( \rho_1 \eta^2 \mathcal{D}_{12} + \rho_2 (1-ik_1 \eta)(1-ik_2 \eta) \right)e^{ik_{12}\eta} (-\eta)^{+i\lambda}\nonumber \\
			&\qquad \qquad \qquad \qquad \times  \underset{f_{k_{3}}(\eta)}{ \underbrace{ \left(N_{f}(-\eta)^{3/2} H^{(2)}_{i\mu}(-k_{3}\eta)\right)}} \nonumber\\
			&= \frac{-i\sqrt{\pi}}{8 k_{1}^3k_{2}^3 k_{3}^{1/2+i\lambda}} \left[ \rho_1 \mathcal{D}_{12} I_{2}\left(-\frac{1}{2}+i\lambda,p\right) +\rho_2 k_{3}^{2} \left\lbrace I_{2}\left(-\frac{5}{2}+i\lambda,p\right) \right. \right. \nonumber \\
			&\left. \left.\hspace{9em} +i p I_{2}\left(-\frac{3}{2}+i\lambda,p\right) - \frac{p^2}{4} I_{2}\left(-\frac{1}{2}+i\lambda,p\right) \right\rbrace \right].
		\end{align}
		where the operator $ \mathcal{D}_{12} $ is,
		\begin{align}
		    \mathcal{D}_{12} = k_{1}^{2} k_{2}^{2} \partial_{k_{12}}^{2} + (- \vec{k}_{1} \cdot \vec{k}_{2}) (1-k_{12} \partial_{k_{12}} + k_{1} k_{2} \partial_{k_{12}}^{2}).
		\end{align}
		It can be simplified in the squeezed limit as,
		\begin{align}
		\mathcal{D}_{12} k_{12}^{\alpha} = \frac{1}{8} (\alpha-1)(\alpha-2) \, k_{12}^{\alpha+2}.
		\end{align}
		We can now put together the full form of $I_{+-/-+}$ diagrams,
		\begin{align}
		    I_{+-}+ I_{-+} = 2 \times \left\lbrace \, I_{k_3}^{(-)}(\lambda,\beta_i^*)\times I_{k_{12}}^{(+)} (\lambda,\rho_i) + \left((\lambda,\beta^*_i,\rho_i) \rightarrow (-\lambda,\beta_i,\rho^*_i)\right) \,\right\rbrace + \text{c.c.} \;,
		\end{align}
		where the integrals at $k_3$ and $k_{12}$ vertices are evaluated in \cref{eq:Ipm3_full,eq:Ipm12_full}, and we have accounted for the symmetry factor.
		
	\subsection{Full analytic calculation of $ I_{++/--} $ diagrams}\label{subsec:Ipp_full}
	As discussed in \cref{Ipp_stat}, the integrals at $k_3$ gets contribution from earlier time which is well separated from the time period that contributes to the integral at $k_{12}$. This takes care of the time-ordering and lets us expand the range of integration to the entire domain. 
	\begin{align}
	     I_{++} &\supset  \,\langle \,  \xi_{0} \xi_{0} \xi_{0}\, \cdot \, \int_{-\infty}^{0} \frac{d\eta_2}{\eta_2^4}(\rho^*_1 (\partial \xi)^2+ \rho^*_2 \xi^2) \, \tchi^{\dagger} \, (-\eta_2)^{-i\lambda} \int_{-\infty}^{0}\frac{d\eta_1}{\eta_1^4}  (\beta_1 \dot{\xi}+ \beta_2 \xi)\, \tchi\, (-\eta_1)^{i\lambda}\, \rangle.
	\end{align}
	A closer look will reveal that $I_{k_3}^{(+)}$ can be obtained by replacing $I_1 \rightarrow I_2$ and $\lambda \rightarrow -\lambda$ with an appropriate change in sign in a few places in \cref{eq:Ipm3_full}.
	\begin{align}\label{eq:Ipp3_full}
	    I_{k_{3}}^{(+)} &= 
			\frac{-i\sqrt{\pi}}{4} \frac{1}{k_{3}^{3/2+i\lambda}}\left[ -\beta_1 I_{2}\left(-\frac{1}{2}+i\lambda,1\right) +\beta_2 \left\lbrace I_{2}\left(-\frac{5}{2}+i\lambda,1\right)+i I_{2}\left(-\frac{3}{2}+i\lambda,1\right) \right\rbrace \right].
	\end{align}
	Similarly, $I_{k_{12}}^{(+)}$ follows from \cref{eq:Ipm12_full} with the replacements $I_2 \rightarrow I_1$, $p \rightarrow -p$, $\lambda \rightarrow -\lambda$, and appropriate sign changes.
	\begin{align}\label{eq:Ipp12_full}
			I_{k_{12}}^{(+)}
			&= \frac{-i\sqrt{\pi}}{8 k_{1}^3 k_{2}^3 k_{3}^{1/2-i\lambda}} \left[ \rho^*_1 \mathcal{D}_{12} I_{1}\left(-\frac{1}{2}-i\lambda,-p\right) +\rho^*_2 k_{3}^{2} \left\lbrace I_{1}\left(-\frac{5}{2}-i\lambda,-p\right) \right. \right. \nonumber \\
			&\left. \left.\hspace{9em} +i p I_{1}\left(-\frac{3}{2}-i\lambda,-p\right) - \frac{p^2}{4} I_{1}\left(-\frac{1}{2}-i\lambda,-p\right) \right\rbrace \right].
		\end{align}
	Then the contribution from $I_{++/--}$ diagrams is
	\begin{align}
	    I_{++}+I_{--} = 2 \times \left\lbrace \, I_{k_3}^{(+)}(\lambda,\beta_i) \times I_{k_{12}}^{(+)}(\lambda,\rho_i^*) +\left((\lambda,\beta_i,\rho_i^*) \rightarrow (-\lambda,\beta^*_i,\rho_i)\right) \, \right\rbrace + \text{c.c.},
	\end{align}
	where now we use expression in \cref{eq:Ipp3_full,eq:Ipp12_full}.
    \bibliographystyle{JHEP}
	\bibliography{refs.bib}
\end{document}